\documentclass{JAIS}

\journal{JAIS-ID}
\vol{2024}

\def\be{\begin{equation}}
\def\ee{\end{equation}}
\def\bea{\begin{eqnarray}}
\def\eea{\end{eqnarray}}

\usepackage{xcolor}

\usepackage{caption}
\usepackage[pagewise]{lineno}
\usepackage{subfigure}
\usepackage{wrapfig}

\usepackage[tight]{units}
\usepackage{amsmath}
\usepackage{amssymb}
\usepackage{graphicx}
\usepackage{fancyhdr}
\usepackage{color}

\usepackage{nicefrac}

\usepackage[utf8]{inputenc}

\articletype{Special Issue}

\usepackage[colorlinks=true,citecolor=blue,urlcolor=blue,linkcolor=blue]{hyperref}
\usepackage{hyperref}

\begin{document}

\title{Scintillating low-temperature calorimeters for direct dark matter search}

\author{M. Kaznacheeva,\auno{1} K. Schäffner,\auno{2}}
\address{$^1$Physik-Department, TUM School of Natural Sciences, Technische Universit\"at M\"unchen, D-85747 Garching, Germany}
\address{$^2$Max-Planck-Institut f\"ur Physik, D-85748 Garching, Germany\\
\vspace{0.2cm}
Corresponding authors: M. Kaznacheeva, K. Schäffner\\
Email addresses: margarita.kaznacheeva@tum.de, karoline.schaeffner@mpp.mpg.de\\
}

\begin{abstract}

The lack of an unambiguous signal for thermally produced dark matter particles in direct detection, indirect detection, and collider searches calls for broadening the search strategies by probing a wider range of dark matter masses with different detection techniques. One of the most common approaches is to search for nuclear recoils induced by dark matter particles scattering off the target material's nuclei. Low-temperature detectors have proven to provide the required performance to probe dark matter masses from $\unit[100]{MeV/c^2}$ to $\unit[100]{GeV/c^2}$ via this channel. Using scintillation light as an ancillary channel is a powerful tool for particle identification and background suppression at the keV-recoil energy scale. The \mbox{CRESST-III} experiment, employing scintillating cryogenic detectors with highly sensitive transition edge sensors and multi-target absorber crystals, achieved unprecedented sensitivities to explore sub-GeV dark matter masses. COSINUS, instead, is a new experiment exploiting the phonon-light technique using sodium iodide crystals with the scope to clarify the long-lasting dark matter claim of the DAMA/LIBRA collaboration. This article reviews the principle of scintillating low-temperature calorimeters with emphasis on the benefits and challenges of this technique for direct dark matter searches in light of the current status and future developments. 

\end{abstract}

\maketitle


\section{Introduction}
\label{s:intro}
Astroparticle physics today is a global research enterprise, and one of its enduring questions is the nature of dark matter (DM)~\cite{appec_report_2022}. The existence of DM is supported by unequivocal evidence from astrophysical observations. Despite comprising over 80\% of the matter in our Universe, DM has so far evaded detection. While the Standard Model of particle physics is very successful in explaining the visible Universe, it has no explanation for DM. The discovery of DM particles and possibly new fundamental forces should revolutionize our knowledge of the fundamental building blocks of nature.

At present, several direct DM searches aim to detect DM particles in earth-bound detectors. 
Since the DM particle mass $m_{\chi}$ remains unknown, performing searches in a wide range of masses is essential. One of the most common approaches to direct detection of DM is measuring the nuclear recoil energy $E_R$ anticipated from DM particles scattering elastically off the detector target material's nuclei. The expected differential recoil rate $\nicefrac{dR}{dE_R}$ of DM particles in a detector's target of the mass $M$ made of a material with the nucleus mass $m_N$ can be described as:

\begin{equation}\label{eq:dRdE}
     \frac{dR}{dE_R}= \frac{M}{m_N} \frac{\rho_{\chi}}{m_{\chi}} \int^{v_{esc}}_{v_{min}(E_R)} dv f(v) v \frac{d \sigma}{d E_R}(v), 
\end{equation}

\noindent where ${\rho_{\chi}}$ is the local DM mass density (typical value for earth-bound detectors is ${\rho_{\chi}=\unit[0.3]{(GeV/c^2)/cm^3}}$~\cite{salucci2010dark}) and $\sigma$ is the interaction cross-section. Under the assumption of the standard DM halo model~\cite{DMhaloModel}, the distribution $f(\vec{v})$ of the DM particle velocities $v$ in the galactic rest frame follows a Maxwell-Boltzmann distribution. In Eq.~\ref{eq:dRdE} it is integrated from the minimal velocity $v_{min}$ able to induce a recoil energy $E_R$ to the galactic escape velocity ${v_{esc}}$ with the typical value of ${v_{esc}=\unit[544]{km/s}}$~\cite{Smith2007,baxter_recommended_2021}. The differential cross-section is an unknown function and can take many forms in different DM models. In the spin-independent (SI) case, the DM particle interacts identically and coherently with protons and neutrons in the target nucleus. Therefore, the cross-section is enhanced in comparison to the DM-nucleon cross-section $\sigma_n$: 

\begin{equation}\label{eq:dSigmadE}
     \frac{d \sigma_{SI}}{dE_R}= \frac{m_N}{2 v^2} \frac{\sigma_n}{\mu_n^2} A^2 F^2(E_R),
\end{equation}

\noindent where $\mu_{n}$ is the DM-nucleon reduced mass and $A$ is the mass number of the target material. The structure of the nucleus is taken into account by introducing the form factor $F(E_R)$, which is relevant for high momentum transfers where the point-like approximation of the nucleus is no longer valid. The most commonly used form factor approximation is the Helm function introduced in~\cite{helm,lewin_smith_1996}. 

If a target nucleus has a non-zero spin, an experiment is also sensitive to spin-dependent (SD) DM-nucleon interactions.  

While theory does not provide precise guidance, there are more and more compelling models of new physics, including those of the dark sector, which naturally provide masses below the weak scale \cite{PhysRevD.79.115016,falkowski_asymmetric_2011,PhysRevD.106.035031,PhysRevD.69.101302}. In this mass regime, low-temperature detectors achieve unprecedented sensitivities thanks to their low thresholds. For DM particles with $m_{\chi} \ll $ the mass of a proton ($\sim\unit[1]{GeV/c^2}$), processes in which the DM scatters inelastically off the nuclei of the detector target material offer interesting perspectives since a significant fraction of the DM’s kinetic energy can be transmitted to the target material. Examples of inelastic processes besides DM scattering with bound electrons (e.g.~\cite{Derenzo_2017,Essig_2021,Catena_2020}), which recently received much attention, are DM scattering with target nuclei (e.g.~Xe and Ar) through the Migdal effect~\cite{Migdal_2022,Migdal_Essig_2023} or scattering accompanied by bremsstrahlung.

\begin{figure}
        \centering \subfigure[]{
        \includegraphics[width=0.55\columnwidth]{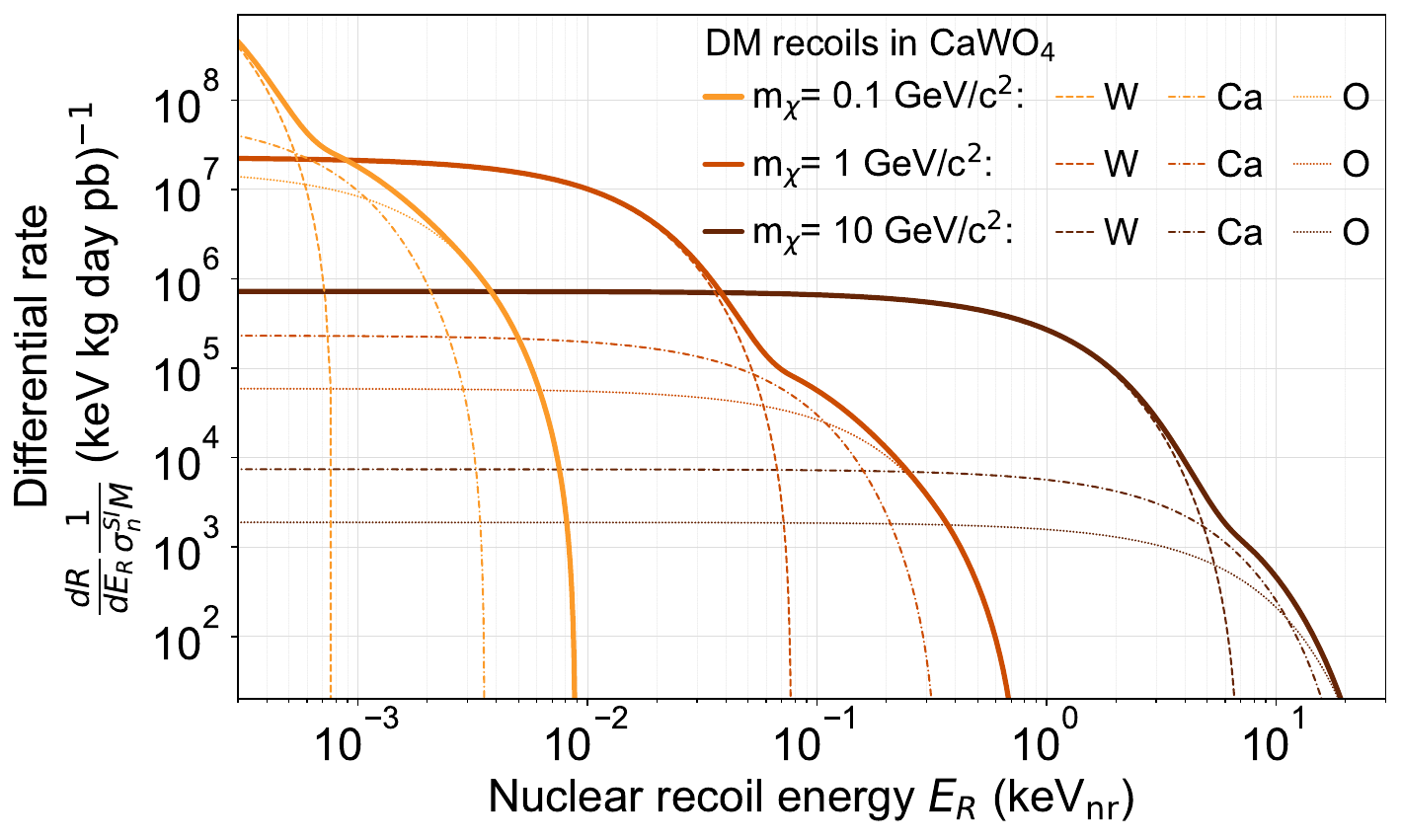}\label{f:Rate}}
        \subfigure[]{
        \includegraphics[width=0.4\columnwidth]{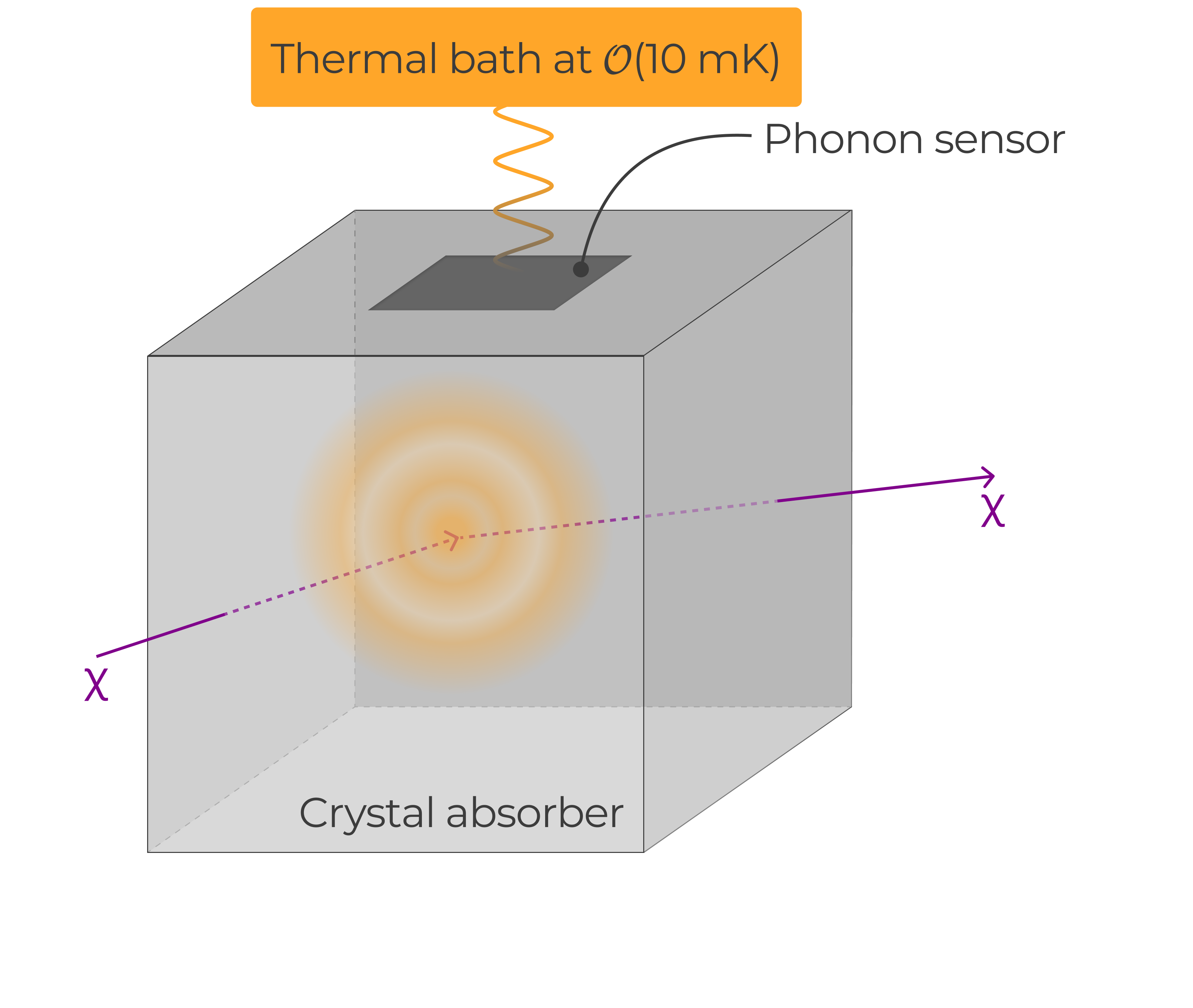}\label{f:CryoDet}}
\caption{(a) Solid lines show the expected nuclear recoil energy spectra for spin-independent interactions of DM particles with masses of  0.1, 1, and ${\unit[10]{GeV/c^2}}$ in a CaWO$_4$ target. Individual contributions from scattering with tungsten, calcium, and oxygen nuclei are shown with dashed, dash-dotted, and dotted lines respectively. (b) Illustration of the general working principle of cryogenic detectors.}
\label{f:DMinCryoDet}
\end{figure}

Since neither the DM mass $m_\chi$ nor the interaction cross-section $\sigma$ is known, these two variables form the parameter space commonly used to express experimental results. Combining information about the expected energy spectrum for different $m_\chi$ and $\sigma$ values with the measured energy spectrum enables the identification of allowed regions within the parameter space in case of a discovery, or the exclusion of certain regions in case of no observed signal.

The differential nuclear recoil spectrum expected from the SI DM-nucleus interaction for several DM particle masses in an ideal CaWO$_4$ detector calculated with Eq.~\ref{eq:dRdE} and~\ref{eq:dSigmadE} is shown in Fig.~\ref{f:Rate}. The rate is normalized to the detector mass $M$ and DM-nucleon cross-section $\sigma_n$. Additionally, individual components of the recoil rate from interactions with tungsten, calcium, and oxygen nuclei are provided for comparison. 
The masses of both the DM particle and the target nuclei determine the kinematics of the scattering process, thus influencing the shape of the expected energy spectrum. The maximum possible recoil energy is defined by ${E_{R}^{max}=\nicefrac{2 \mu_N^2 v^2}{m_N}}$, where $\mu_N$ is the DM-nucleus reduced mass. The lower the mass of the DM particle, the lower is the maximum possible recoil energy. This establishes a strict requirement for the detection threshold in DM search experiments. For heavy tungsten nuclei, the $A^2$-enhancement of the cross-section from Eq.~\ref{eq:dSigmadE} results in higher expected event rates. However, lighter target nuclei, such as oxygen, extend the energy spectrum to higher, and therefore more easily accessible, values.

\begin{figure}
        \centering \includegraphics[width=1\columnwidth]{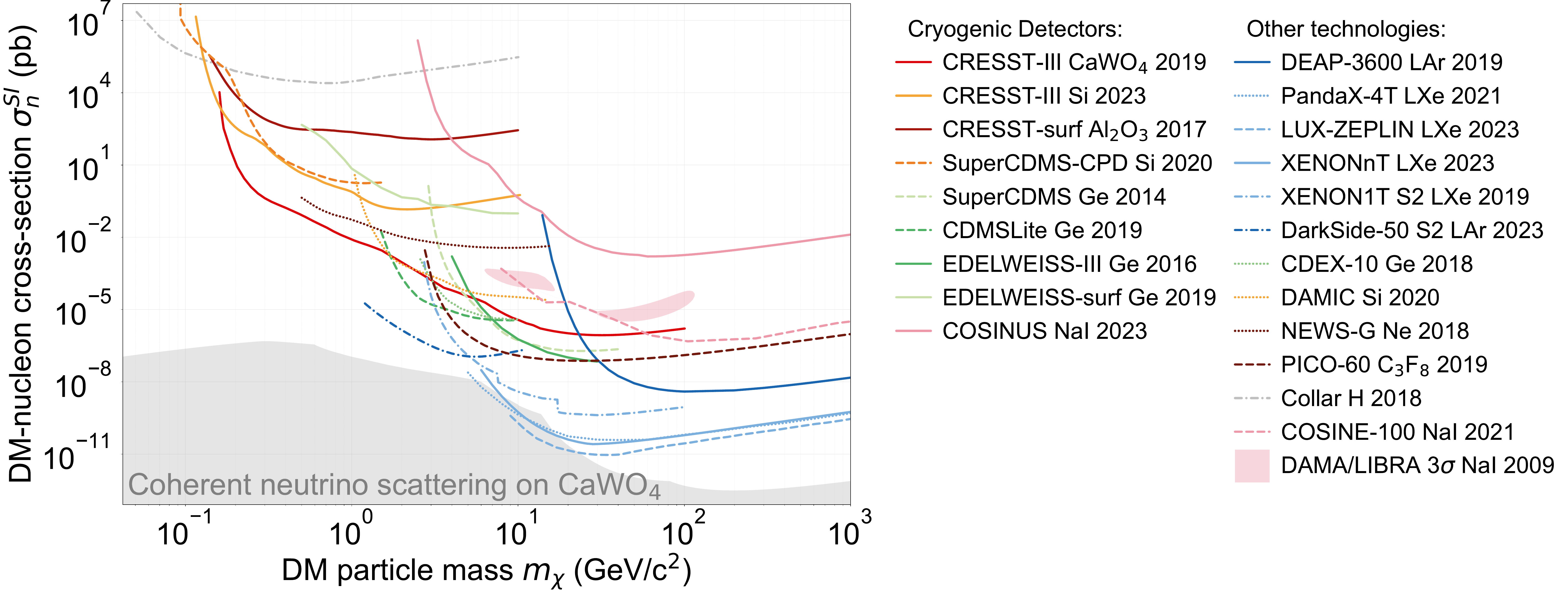}
\caption{Results of DM searches via spin-independent elastic DM-nucleus scattering obtained with various detection principles and target materials indicated in the legend. The regions above the lines are excluded. The following results are shown. With cryogenic detectors in deep-underground setups: {CRESST-III CaWO$_4$~\cite{detA}}, {CRESST-III Si~\cite{CRESST_SiDet}}, {SuperCDMS~\cite{supercdms_DM_2014}}, {CDMSLite~\cite{CDMSlite_DM_2019}}, {EDELWEISS-III~\cite{EdelweissIII2016}}, {COSINUS~\cite{cosinus_collaboration_deep-underground_2023}}; and in the surface facilities: {CRESST-surf~\cite{CRESSTsurf2017}}, {SuperCDMS-CPD~\cite{CPD2021DM}}, and EDELWEISS-surf~\cite{edelweissSurf2019}. With liquid noble gas experiments: {DEAP-3600~\cite{deap3600_DM_2019}}, {PandaX-4T~\cite{pandax-4t_DM_2021}}, {LUX-ZEPLIN~\cite{LZ_DM_2023}}, {XENONnT~\cite{xenonNt_DM_2023}}; with ionization signal only (S2): {XENON1T S2~\cite{xenon1t_S2_2019}}, {DarkSide-50 S2~\cite{darkside50_S2_2023}}. With $p$-type point contact Ge detectors {CDEX-10~\cite{cdex_DM_limits_2018}}, with CCD sensors {DAMIC~\cite{damic_DM_2020}}; with spherical gaseous proportional counter {NEWS-G~\cite{news-g_DM_2018}}; with the bubble chamber {PICO-60~\cite{pico60_DM_2019}}; with hydrogenated organic scintillators by {J.~I.~Collar~\cite{collar_DM_2018}}; with NaI: {COSINE-100~\cite{cosine100_DM_2021}} and contours compatible with {DAMA/LIBRA} results~\cite{dama/libra_2008} calculated in~\cite{Savage_dama_islands_2009}. The gray area shows the parameter space wherein the background, induced by coherent scattering of solar and atmospheric neutrinos off the target nuclei, appears in a DM search experiment using CaWO$_4$ absorbers calculated in~\cite{nufloor_cawo4_2015,fuchs_phd2023}. It is commonly referred to as the neutrino floor or, more recently, neutrino fog~\cite{nufog_2021}.}
\label{f:limits}
\end{figure}

For the classical weakly interacting massive particle (WIMP) mass regime (${m_{\chi}\gtrsim\unit[10]{GeV/c^2}}$), the necessary detector thresholds are relatively easy to achieve with current technologies; building large detectors to ensure sufficient exposure to explore the lower DM cross-sections becomes the main challenge. Indeed, liquid noble gas detectors with multi-tonne-year exposures are leading the field for both SI and SD  interactions in this mass range thanks to their extremely low internal backgrounds and ease of scalability. Instead, cryogenic detectors have consolidated in the past years their pioneering role in the low-mass regime (${\unit[100]{MeV/c^2}\lesssim m_{\chi}\lesssim\unit[1]{GeV/c^2}}$) as they combine both very high sensitivity and a multi-target approach.
In fact, due to the higher expected event rate of lighter DM particles, gram-scale detectors like cryogenic scintillating crystal calorimeters, which achieve ultra-low energy thresholds ${\mathcal{O}(\unit[1]{eV})}$, are sufficient to explore new parameter space with unprecedented sensitivity. Here, further decreasing of detector thresholds poses the primary goal. These effects can be seen from the current landscape of experimental results for DM exclusion limits obtained via SI elastic DM-nucleus scattering shown in Fig.~\ref{f:limits}. We aimed to include the most stringent constraints obtained to-date with various target materials and distinctive detection technologies.  

While the general terms sub-GeV or light DM can cover many orders of magnitude of DM particle masses, in this review we use them to refer to the mass range from about $\unit[100]{MeV/c^2}$ to about $\unit[1]{GeV/c^2}$. The lower limit of this range depends strongly on the detector threshold and is, therefore, expected to be extended to lower values in the coming years.

As is demonstrated in Fig.~\ref{f:limits}, low-temperature detectors are able to sense nuclear recoil signals to search for sub-GeV DM particles. Since similar requirements are necessary for precision  measurements of coherent elastic neutrino-nucleus scattering (CE$\nu$NS)~\cite{cevns_1984}, they are today also employed by various experiments~\cite{nucleus_2019,ricochet_status_2023,resnova_2020,miner_2017,neutrino_snowmass_2022} to study fundamental neutrino physics. Furthermore low-temperature detectors are also key players in experiments~\cite{cuore_2022,cupid0_final_2022,cupid-mo_final_2022,amore-I_status_2022,0nubb_review_agostini_2023} searching for neutrinoless double-beta decay~\cite{furry_0nubb_1939}. 
 
\section{Low-temperature Calorimeters for Dark Matter Searches}
    
\label{LTDs}
\subsection{The Concept}
A particle interaction in the target, such as a nuclear recoil, leads to an energy deposition that can cause different types of excitation processes, e.g. ionization, scintillation, and heat, depending on the material's properties. In crystalline detectors, a dominant part of the deposited energy is converted to vibrations of the crystal lattice, i.e. phonon excitations, that lead to a temperature rise $\Delta T$ that can be measured.

From the third law of thermodynamics, the energy deposition $\Delta E$ can be defined precisely from the temperature increase: ${\Delta T=\nicefrac{\Delta E}{C}}$. For dielectric crystals, the heat capacity $C$ is dramatically reduced with the temperature as ${C \propto (T/\Theta_{D})^3}$, when ${T \ll  \Theta_{D}}$, the Debye temperature of the material. Therefore, to ensure a large enough and measurable temperature rise, a small heat capacity and, thus, low operating temperatures are required. To enhance the sensitivity, detectors based on cryogenic calorimetry are operated at temperatures of ${\mathcal{O}(\unit[10]{mK})}$ using wet or dry $^4$He/$^3$He-dilution refrigerators\footnote{Dilution refrigerators are machines that provide continuous cooling and operation conditions at $\unit[<10]{mK}$. While wet dilution refrigerators rely on liquid nitrogen and liquid helium for the pre-cooling to \unit[4]{K}, the new dry refrigerators employ pulse-tube coolers and make the need for liquid nitrogen and liquid helium obsolete. Instead, for both types, the cooling power at the mK-stage is provided by dilution of the mixture of the $^4$He and $^3$He isotopes.}.

A typical cryogenic detector is schematically shown in Fig.~\ref{f:CryoDet}. When a particle interaction occurs within the crystal absorber's volume, first, the energy deposition creates optical phonons that almost instantaneously decay into athermal phonons. Athermal phonons ballistically propagate through the absorber's medium until they eventually thermalize via inelastic scattering off the crystal surfaces. The detector then returns to equilibrium via a weak coupling to the thermal bath. Depending on the chosen technology, the sensor attached to the crystal can measure both athermal and thermal phonon signals.

\subsection{The Phonon Sensors}
\label{s:sensors}
The sensor is a critical detector component that converts a temperature rise to a readable signal. Next, we describe different types of phonon sensors used for macroscopic cryogenic detectors in DM search experiments. 

\textbf{Neutron-transmutation-doped (NTD)} sensors are thermal sensors made of semiconducting crystals doped using intense neutron irradiation~\cite{haller1994}. This leads to a highly homogeneous dopant concentration inside the crystal. Its resistance depends strongly on the temperature and therefore allows measuring the heat signal through a drop of the NTD voltage bias~\cite{Wang1990}. HPGe crystals equipped with Ge-NTD sensors are used for DM search by the EDELWEISS experiment~\cite{EdelweissIII2016,edelweissSurf2019,edelweissElectrons2020}. They apply an electric field to a detector and read out an ionization signal simultaneously with the heat channel. Depending on the regime (strength of the voltage bias applied), this enables particle identification or enhancement of the phonon signal by the Neganov-Trofimov-Luke (NTL) amplification process~\cite{Neganov1985,Luke1988}.

Another type of the sensor is the \textbf{Kinetic Inductance Detector (KID)}~\cite{day2003KID}. Using KIDs as phonon sensors for massive particle detectors in rare-event searches was proposed in~\cite{Swenson2010KIDs} and~\cite{Moore2012KIDs}. A KID forms a resonator whose frequency depends on the Cooper pair density in the material. When athermal phonons from a particle interaction in the absorber crystal reach a superconducting KID sensor, they can break Cooper pairs and change the kinetic inductance of the sensor. This leads to a measurable frequency shift of the resonance. Technically, a KID detector consists of a thin-film superconducting (e.g. Al) meandered kinetic inductor in parallel with an interdigitated capacitor. The resonance frequency of a KID is adjustable, and many of them can be coupled to the same feedline. This provides the main advantage of the KIDs for DM search experiments - their multiplexing capability. Arrays of KID sensors for DM or CE$\nu$NS experiments are currently under development, e.g. in~\cite{wen2022KIDs} and~\cite{colas2023KIDs}, and are implemented in the recent BULLKID project aiming for a kg-scale low-threshold detector~\cite{BULLKID2022}.

\begin{figure}
        \centering \subfigure[]{
         \includegraphics[width=0.49\columnwidth]{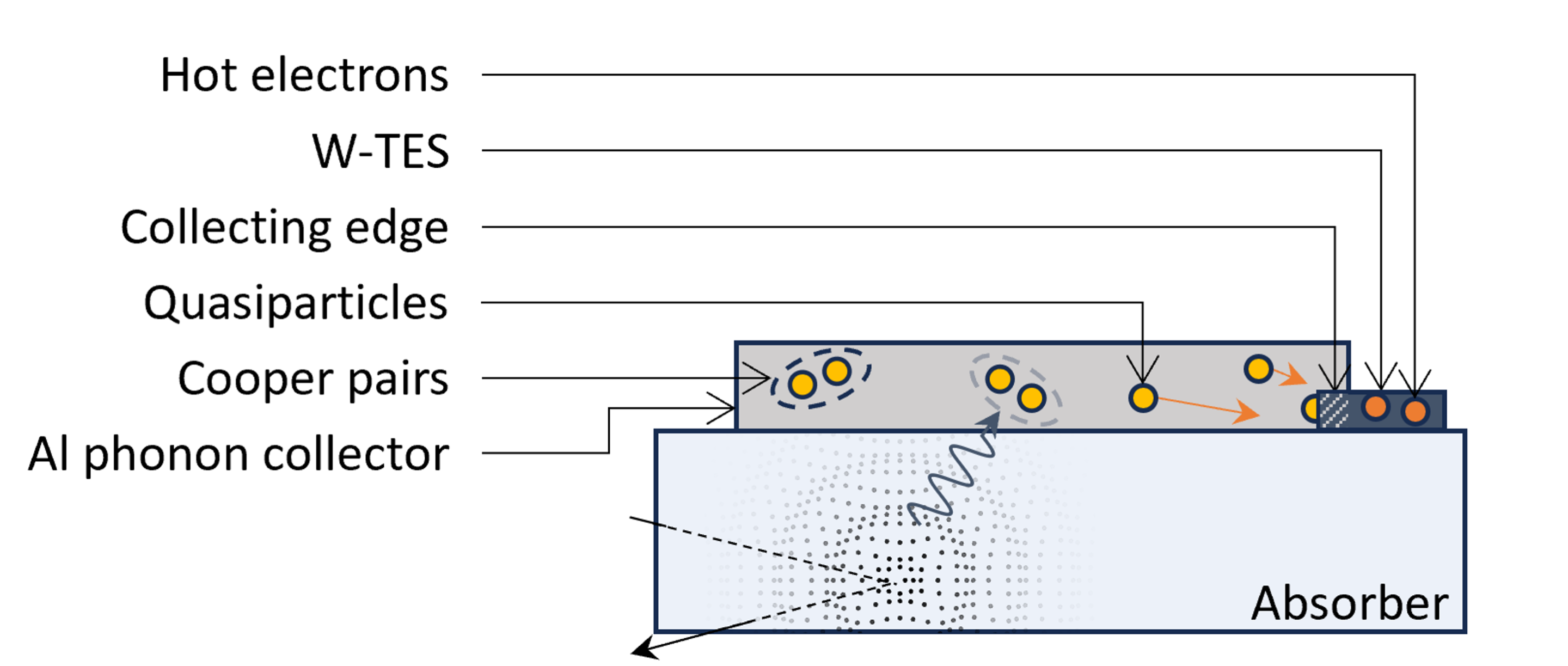}\label{f:PC}}
        \subfigure[]{
        \includegraphics[width=0.49\columnwidth]{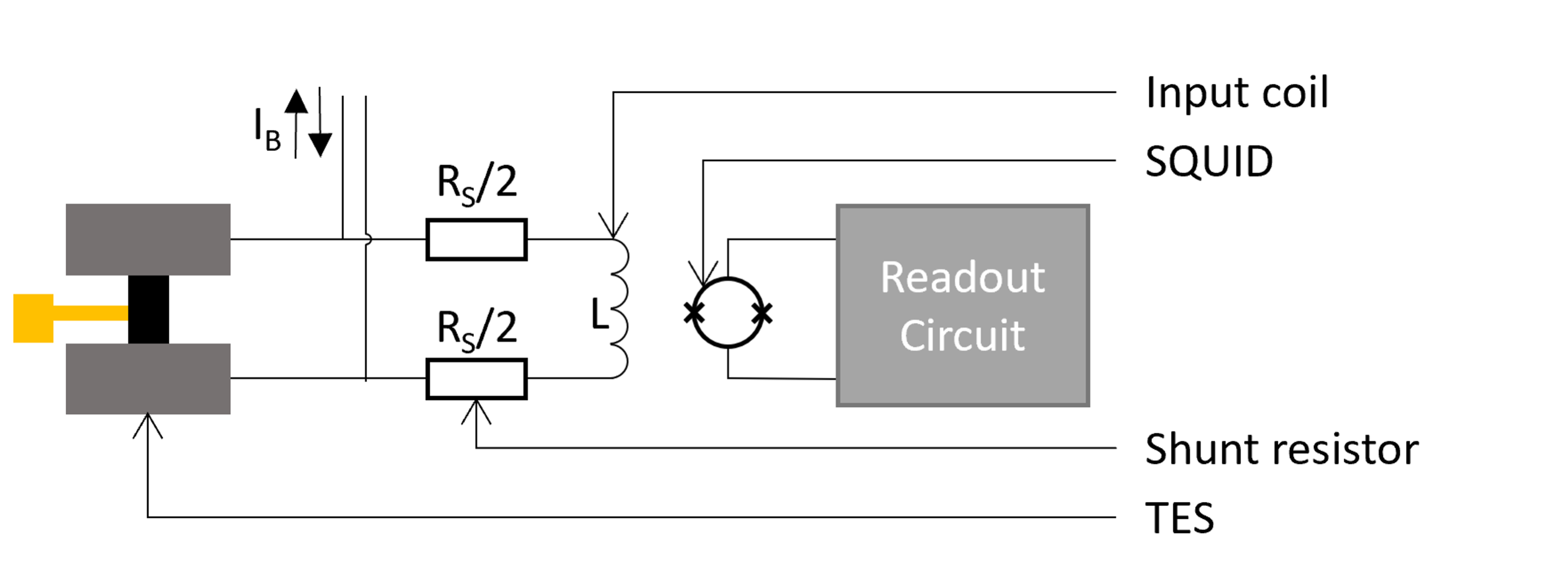}\label{f:SQUIDReadOut}}
\caption{(a) Schematic representation of the signal formation process in a transition edge sensor with phonon collectors or a QET. (b) Readout scheme of a transition edge sensors relying on a SQUID-based amplification.}
\label{f:REadOut}
\end{figure}

Among the most sensitive sensors at low energies currently used for DM searches with cryogenic particle detectors is the \textbf{Transition Edge Sensor (TES)}. It consists of a thin metal film operated at temperatures in between its super- and normal-conducting state. Since the transition curve is very steep, any small temperature rise results in a significant TES resistance change. 

A common material for TES is tungsten. In its $\alpha$ crystalline phase, tungsten has a typical transition temperature $T_c$ as low as \unit[15]{mK}. To enhance the signal, larger coverage of the absorber crystal's surface is beneficial. To increase the area of the sensor without increasing the heat capacity, a TES can be fabricated with a small overlap with another superconducting material with a $T_c$ higher than the TES' $T_c$, such as aluminum. A particle interaction in the crystal creates high-frequency phonons with energies of a few meV, so-called non-thermal phonons, as thermal energies at $\unit[10]{mK}$ correspond to only about \unit[1]{$\mu$eV}. This initial population decays quickly to a stable distribution with a mean frequency of about \unit[100]{GHz}~\cite{probst_model_1995} and allows the phonons to spread ballistically over the crystals' volume. Once absorbed in the aluminum phonon collectors, they break Cooper pairs and produce excited quasi-particles. This process goes on until the energy of the phonons drops below twice the band gap value. The phonons with smaller energies get re-emitted into the crystals and are responsible for the slow component of the measured signal. Instead, the quasiparticles in the Al randomly diffuse and, with a certain probability, reach an overlap region before recombining. If a quasiparticle reaches an edge, where the band gap is zero, it transfers its energy to the electron system of the W-TES, thereby heating it. This comprises the fast and dominant component of the signal. Fig.~\ref{f:PC} shows a schematic illustration of the signal formation following those steps. This approach, known as Quasiparticle-trap-assisted Electrothermal-feedback TES (QET)~\cite{Irwin1995QET} or Phonon Collectors (PCs)~\cite{Ferger1996PhCollectors}, is a powerful tool to enhance the athermal signal. 

Low-impedance TES, like W-TES, require a readout based on a Superconducting Quantum Interference Device (SQUID)~\cite{superconductor_electronics_2019,SQUID_Handbook_2004}. A typical SQUID-based read-out circuit is simplistically shown in Fig.~\ref{f:SQUIDReadOut}. The bias current $I_B$ is split into two branches between the TES and a shunt resistor $R_S$. The current flowing through one of the branches (in the shunt branch for the example in Fig.~\ref{f:SQUIDReadOut}) is coupled into the SQUID via an input coil $L$. Changes in current flow caused by the TES resistance changes lead to alterations in the magnetic field of the coil, to which the SQUID is sensitive. A current change related to the energy deposition to the absorber can be measured by reading out the current through the input coil.

The W-TES with Al PCs attached to crystal targets of different materials (typically CaWO$_4$) is used in the CRESST DM search experiment~\cite{detA} presented in Sec.~\ref{s:cresst}. In case of a scintillating crystal, a simultaneously measured light signal employing a separate cryogenic light absorber read out by TES is used for particle identification, as discussed in Sec.~\ref{PID}. 

Ge or Si targets equipped by W-QET arrays are operated by the SuperCDMS collaboration~\cite{CDMS_QET_2006,SuperCDMS_atSoudan_detectors_2013,CDMSII_first_2013,CDMSlite_DM_2019,SuperCDMS2017Projections,SuperCDMS_snowmass_strategy_2023} and related low-threshold projects: SuperCDMS-HVeV~\cite{HVeV2021Performance,HVeV_DM_2018,HVeV2021DM} and SuperCDMS-CPD~\cite{CPD2021Performance,CPD2021DM}. Similarly to EDELWEISS, if a voltage bias is applied to the crystal, particle discrimination or NTL amplification can be achieved. 

The EDELWEISS experiment, usually using thermal NTD sensors, has recently operated a high-impedance NbSi TES, which, in contrast to W-TES, relies on conventional amplifiers. Exploiting sensitivity to athermal phonons allowed achieving lower energy thresholds with a massive \unit[200]{g} Ge detector~\cite{edelweiss2023NbSiTES,edelweiss2022NbSiDM}.

The COSINUS DM search experiment presented in Sec.~\ref{s:cosinus} uses a novel remote TES (\textbf{remoTES}) approach to read a phonon signal from absorber crystals, such as NaI, whose properties do not allow TES fabrication directly on their surfaces. First proposed in~\cite{pyle2015remoTES}, the remoTES design was realized within the COSINUS collaboration's effort~\cite{COSINUS2023FirstRemoTES}. In this design, a W-TES is fabricated on an Al$_2$O$_3$ wafer separated from the absorber crystal. The absorber crystal has a gold pad coupled to the TES via a gold wire. Thus the remoTES design allows to avoid the use of an intermediate glued carrier crystal used previously, e.g.~\cite{CRESST2009Composite}, and enables to employ a variety of new target materials. The remoTES layout also provides ease in fabrication, an advantage in particular for larger detector arrays. Besides COSINUS,  RICOCHET, an experiment for precision measurements of CE$\nu$NS, is considering using this detector layout \cite{RICOCHET}, and AMORE \cite{AMORE_2019} utilizes thin gold films as phonon absorbers connected to metallic magnetic calorimeters (MMCs) to search for the neutrinoless double-beta decay.

Under the umbrella TESSERACT~\cite{tesseract_snowmass_letter_2021,snowmass2021} (Transition Edge Sensors with Sub-eV Resolution And Cryogenic Targets), there are also new players in the field of cryogenic calorimeters. They aim to even further extend the DM search region to the sub-eV scale for both DM-nucleus scattering and DM-electron interactions by using solid absorbers (Al$_2$O$_3$, SiO$_2$, and GaAs) within the SPICE experiment and superfluid $^4$He as a target in the HeRALD experiment~\cite{HeRALD_demo_2023}. Another new superfluid helium target project is DELight~\cite{delight_2023}. The signal, in the latter, will be collected by a crystal wafer equipped with an array of MMCs~\cite{mmc_1993,mmc_2005}. In a  MMC the change in magnetization of a paramagnetic material in a weak magnetic field is used as a measure of the energy deposition in the absorber. Recently, an unprecedented energy resolution of $\Delta E_{\text{FWHM}}=\unit[1.25]{eV}$ at \unit[5.9]{keV} has been demonstrated for a tiny absorber with a mass of $\sim$\unit[1]{$\mu$g}~\cite{mmc_2024}.

\begin{figure}
        \centering \includegraphics[width=1\columnwidth]{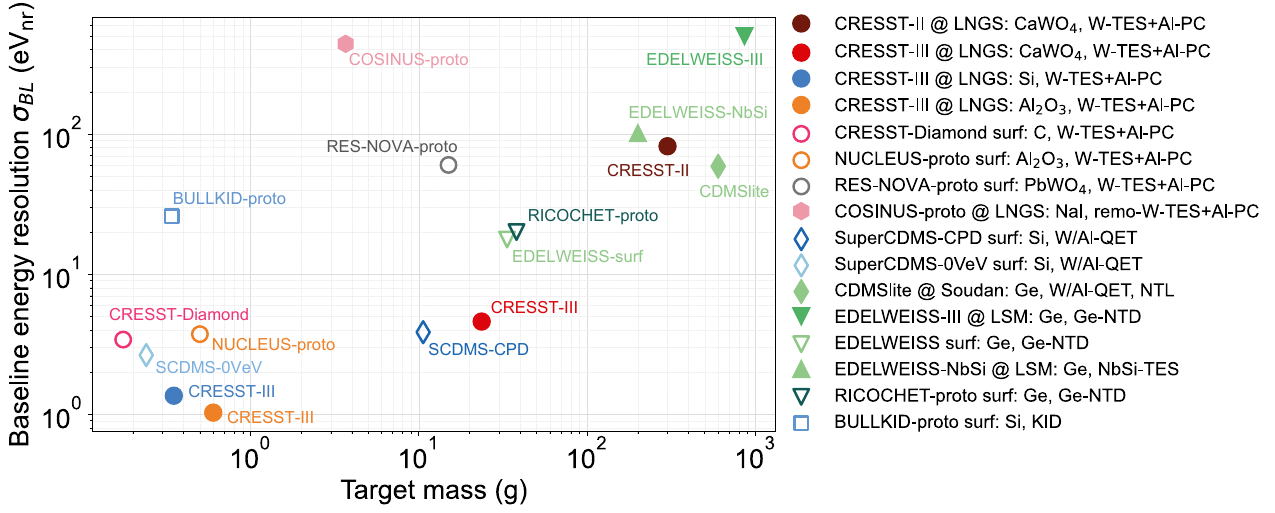}
\caption{Overview of the baseline resolution  values $\sigma_{BL}$ achieved in recent macroscopic cryogenic rare event search experiments and prototypes employing different target materials and phonon sensor technologies whose sensitivity and physics goals rely on low energy thresholds: {CRESST-II}~\cite{CRESST-II_2016results}, {CRESST-III} CaWO$_4$~\cite{detA}, {CRESST-III} Si~\cite{CRESST_SiDet}, {CRESST-III} Al$_2$O$_3$~\cite{cresst_singlephotons_2024}, {CRESST-Diamond}~\cite{diamond2022,diamond2023DM}, NUCLEUS prototype~\cite{nucleus1g_2017,CRESSTsurf2017}, {RES-NOVA} prototype~\cite{resnova2022operation}, COSINUS-prototype~\cite{cosinus_collaboration_deep-underground_2023}, {SuperCDMS-CPD}~\cite{CPD2021Performance,CPD2021DM}, {SuperCDMS-0VeV}~\cite{HVeV2021Performance}, {CDMSlite~\cite{cdmslite_run2_2018}}, {EDELWEISS-III~\cite{edelweissIII_performance_2017,EdelweissIII2016}}, {EDELWEISS-surf}~\cite{edelweissSurf2019}, {EDELWEISS-NbSi}~\cite{edelweiss2022NbSiDM}, {RICOCHET} prototype~\cite{ricochet_30eVee_2023}, {BULLKID} prototype~\cite{BULLKID2022}. Similar colors denote common target materials, while distinctive marker shapes represent different phonon sensor technologies. The filled markers show the measurements carried out in underground facilities, specified in the legend. Measurements taken in above-ground facilities are displayed with the open markers and marked with \textquotesingle surf\textquotesingle~in the legend. Smaller target masses within a specific sensor technology result in lower baseline resolution values, which correspond to higher detector sensitivity. COSINUS drops out of line here since using a non-standard target material, sodium iodide (NaI), which is less suitable for low-temperature detectors due to its soft and hygroscopic nature and its higher heat capacity.}
\label{f:BLvsM}
\end{figure}

Optimization of a cryogenic detector is a complex and multi-parameter task. However, it is generally easier to achieve a lower energy threshold with a smaller absorber crystal \cite{probst_model_1995,nucleus1g_2017}. Since, as was discussed in Sec.~\ref{s:intro}, lowering the energy threshold allows probing lower DM masses, many cryogenic DM searches in recent years have reduced their target masses to ${\mathcal{O}(\unit[10]{g})}$ and below. While the detection threshold defines the sensitivity to light DM, it is largely a matter of choice, and there is currently no commonly accepted convention on how to make this choice. However, the threshold is directly related to the baseline energy resolution, i.e. the resolution at \unit[0]{eV} energy deposition, defined by finite baseline noise, which can be seen as a more fundamental parameter for comparing detector performance of different experiments. 
Fig.~\ref{f:BLvsM} shows the relationship between the crystal absorber masses and the corresponding baseline energy resolutions $\sigma_{BL}$ for nuclear recoils with macroscopic cryogenic detectors featuring different phonon sensors and target materials. The selection criteria for the depicted results are: massive gram-scale low-temperature detectors, both prototypes and existing experiments in the field of rare events search, published in recent years and particularly focusing on searches that pushed their sensitivity for low thresholds. Detectors with smaller masses are expected to achieve higher sensitivity due to their reduced heat capacity. Furthermore, a higher heat capacity at the same mass will worsen the achievable performance for the same argument. Note that an adaptation of the sensor is also required for consistent and predictable improvements. Since the designs of the new detector generations, which aimed to push the thresholds to lower values, had been developed with these arguments in mind, the trend towards lower baseline resolution values for smaller target masses within the same sensor technology is prominent. Notably, measurements conducted on the surface may be influenced by instabilities and high particle rates, reducing sensitivities compared to measurements carried out in low-background facilities.

\section{Scintillating Crystals for Particle Identification}

\subsection{Scintillator Principle and Material Choice}
A scintillating crystal is known as a target material that can emit light (IR to UV) following the absorption of energy due to e.g.~a particle interaction. Nowadays scintillators operated at room temperature encompass a wide field of applications ranging from particle detectors in fundamental physics to medical imaging and homeland security usage. 
Employing scintillating materials at milliKelvin temperature for cryogenic calorimeters is attractive as they offer a second detection channel: a fraction of the deposited energy in the crystal is emitted in the form of scintillation light. The simultaneous detection of both the phonon signal in the crystal and the scintillation signal, by using a separate cryogenic light absorber, is a powerful tool for particle identification on an event-by-event basis. Indeed, since the dominant background in cryogenic detectors is constituted by $\beta$/$\gamma$-particles which produce more scintillation light at the same deposited energy than the sought-for nuclear recoils, the phonon-light technique is the key enabler for background identification and rejection; this detection approach was first proposed almost 35 years ago~\cite{GONZALEZMESTRES_First_Scint_Cryo_Det_1989}. 

The scintillation process is a sequence of the following steps: (i) creation of primary electron-hole (e-h) pairs by primary carriers such as excited electrons or a recoiling nucleus from a particle interaction in the crystal (ii) relaxation of initial excitations by the creation of secondary electrons, holes, ion collision cascades and other excitations (iii) thermalization into e-h pairs around the band gap energy (iv) energy transfer from the electrons/holes/e-h pairs to the luminescence centers and their excitonic states and, (v) emission from the luminescence centres~\cite{Rodnyi_scintillation_1995}. 

The list of scintillating materials is very long and the choice of the scintillator(s) needs to be aligned to the requirements of the experimental search strategy. For rare event searches operated at milliKelvin temperatures, inorganic scintillators turned out to be a suitable choice as they combine: (i) a high light output of $ \mathcal{O}(\text{1\%})$ combined with excellent optical quality, (ii) single crystal quality which is crucial to ensure good phonon propagation, (iii) chemical purity to suppress internal radioactive contaminations which could mimic the sought-for signals, and (iv) established crystal growing processes including purification and thermal treatment of the starting materials and as-grown ingots, respectively.

Commonly used in the field of direct DM detection are tungstates (CaWO$_4$, CdWO$_4$, ZnWO$_4$, and PbWO$_4$), oxide scintillators with Al$_2$O$_3$ and LiAlO$_2$ as well as alkali metal iodides (CsI and NaI). Molybdates, oxides as TeO$_2$\footnote{Not a scintillating crystal.} and selenides reveal also excellent properties and find applications in the field of the search for the neutrinoless double-beta decay~\cite{0nubb_review_agostini_2023} as well as CE$\nu$NS. The interested reader is referred to~\cite{Poda_review_scintillators_2021} for a detailed review on scintillators for rare event searches. 

\subsection{Light Quenching and Particle Identification}
\label{PID}

The different energy loss mechanisms of particles interacting in a scintillator are responsible for the dependence of the scintillation light output on the particle type. If the energy loss $\nicefrac{dE}{dx}$ of the particle is large in comparison to the spacing of the luminescence centers, saturation hinders the overall light output. Instead, for low ionizing particles like electrons, where the interaction distance is large in comparison to the distance of luminescence centers, the saturation effect is negligible. 
This observation and mechanism is described by a semi-empirical relation proposed by Birks~\cite{Birks_paper_1951,Birks_theory_1964_book} 

\begin{equation}\label{eq:Birks}
     \frac{dL}{dx}= S \cdot \frac{\nicefrac{dE}{dx}}{1+k \cdot B \cdot \nicefrac{dE}{dx}},
\end{equation}

\noindent where $\nicefrac{dL}{dx}$ is the differential scintillation light yield ($LY$) per path length $x$, $S$ is the scintillation efficiency, $\nicefrac{dE}{dx}$ is the particle-specific energy loss, $B\cdot\nicefrac{dE}{dx}$ is the density of excitation centers along the particles' path length, and $k$ is a quenching parameter. Commonly the product $k\cdot B$ is treated as a single parameter called Birks' factor which depends on the material~\cite{Birks_theory_1964_book,TRETYAK_OF_2010}.

The term quenching factor (QF) is used in order to describe the effect of the interaction mechanism of the particles on the scintillation efficiency of the scintillator. In the field of direct DM detection, the QF is usually defined as the ratio of the light response for electrons to the one of other particles ($\alpha$s, neutrons, ...) at the same energy. 

\begin{figure}
        \centering 
        \includegraphics[width=1\columnwidth]{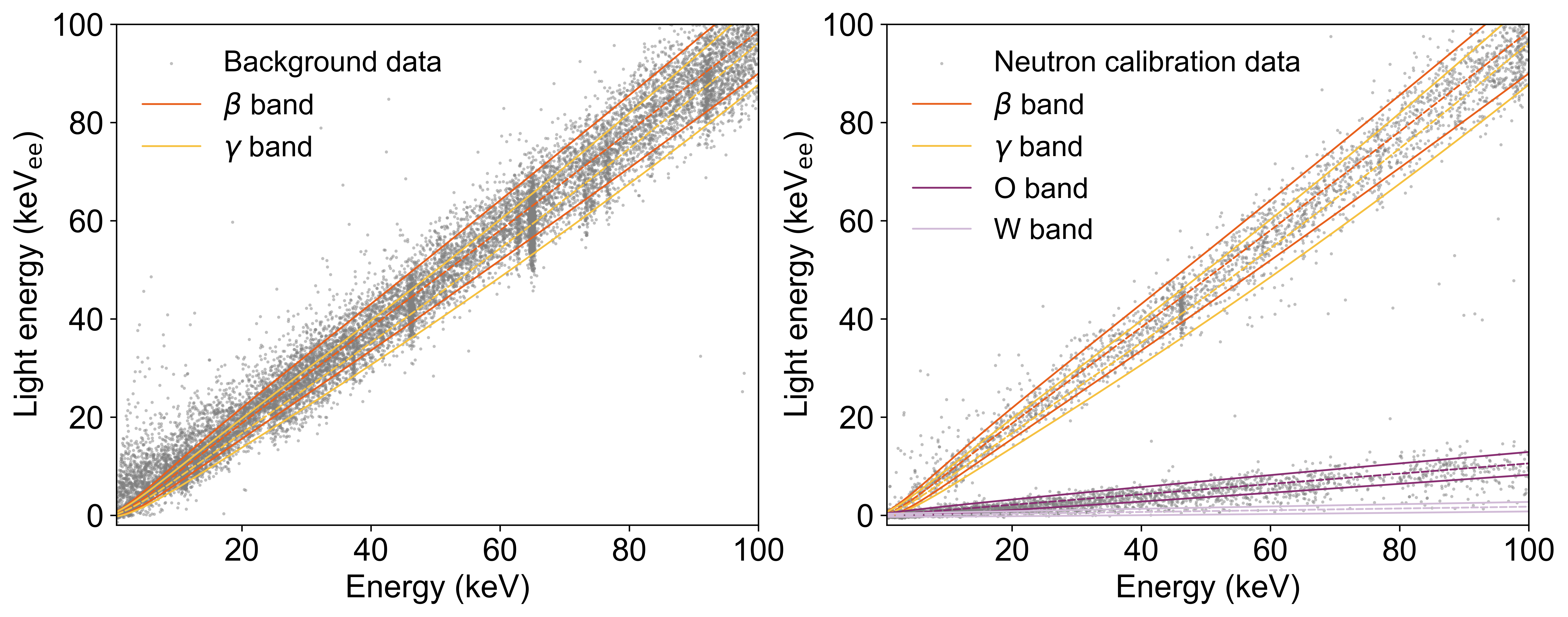}
\caption{Scatter plot of the energy in the light detector in keV$_\text{ee}$ vs.~the total deposited energy in keV of the TUM40 CRESST-II detector module~\cite{Strauss_2015}. The total energy was calculated by considering the measured energies in both the phonon and light channels and the scintillation efficiency of the crystal, as described in~\cite{CRESSTII2014}. Left: Background data for a CaWO$_4$ crystal showing the populated $\beta$/$\gamma$-bands from intrinsic backgrounds of the crystal and its surroundings.  Right: Adding an AmBe neutron source results in additional bands with a smaller slope due to O, Ca, and W recoils. A clear separation of $\beta$/$\gamma$-events from nuclear recoil events is achieved. To a certain extent, even recoils from O can be separated from W. The event population above the $\beta$/$\gamma$-bands, exhibiting a higher light signal, results from electrons that, before being absorbed by the crystal, produce scintillation light in the foil surrounding it.}
\label{f:PID_CRESST}
\end{figure}

The potential of the light-phonon technique regarding particle identification on an event-by-event base is demonstrated in Fig.~\ref{f:PID_CRESST}. The emitted scintillation light by highly ionizing particles like $\alpha$s and particles interacting with target nuclei, like neutrons, is hindered/quenched in comparison to the one induced by $\beta$/$\gamma$-interactions. The energy of a gamma particle is deposited in a crystal by first being transferred to multiple electrons. However, in this process, the fraction of the initial energy emitted in the form of scintillation light is smaller compared to that emitted by an electron of the same energy. As a result, the $\gamma$ band appears slightly lower than the $\beta$ band. \cite{cresst_likelihood_2024}.
The different event classes appear in separated and distinct bands characterized by different slopes in the light vs.~energy plane. The discrimination power, in particular among different recoiling nuclei in a multi-element target as CaWO$_4$ (see Fig.~\ref{f:PID_CRESST}) depends on the overall scintillation efficiency and the sensitivity of the cryogenic light detector as discussed in Sec.~\ref{LDs}.

To be complete, particle discrimination can be also achieved using semiconductor-based absorbers like Si, Ge~\cite{EdelweissIII2016,supercdms_DM_2014}. Here, instead of the scintillation light, the ionization signal, read together with the phonon signal, is used to separate $\beta$/$\gamma$-events from nuclear recoils. 
Cryogenic semiconductor calorimeters are a competing technology for light DM search. Much like the scintillating low-temperature calorimeters, they are characterized by excellent low energy thresholds and e.g.~the SuperCDMS experiment, which is presently in commissioning phase at the SNOLAB facility, has a broad physics program. Searches include nucleon-coupled DM with small and medium-size detectors as well as looking for interactions of DM with electrons to investigate axion and axion-like particle DM and dark photon DM~\cite{supercdmscollaboration2023strategy}.

\subsection{Light Absorbers and Temperature Sensors}
\label{LDs}
The material used for the fabrication of cryogenic light detectors are pure Si and Ge, and so-called silicon-on-sapphire (SOS) wafers which consist of Al$_2$O$_3$ wafers with a thin layer of epitaxially grown Si of \unit[1]{$\mu$m} thickness. Typical dimensions of the light absorbers used e.g.~in CRESST-III are \unit[20x20x0.4]{mm${^3}$}.
The wavelength of the emitted scintillation light dictates the choice of the light absorber material. 
For direct DM searches mainly SOS wafers are used to collect the blueish light emitted from the crystal scintillators like CaWO$_4$ and LiAlO$_2$. Besides the wavelength argument, also the fabrication of the TESs is well established on Al$_2$O$_3$ and thus allows to produce very sensitive sensors. The TESs are optimized for the detection of light: the overall size of both the tungsten film and the phonon collectors is adapted in order to achieve eV-scale baseline resolutions. Recently CRESST-III presented results of a SOS light absorber with an energy baseline resolution of \unit[1.0]{eV}: single photon peaks from interactions in a Al$_2$O$_3$ absorber crystal could be detected. This is the first ever achieved single-photon sensitivity using this technique in CRESST~\cite{cresst_singlephotons_2024}. 

Also more massive light absorbers in the shape of a beaker made from high-purity silicon (\unit[40]{mm} in diameter and height, \unit[0.6]{mm} wall thickness) are used. Regardless of their elevated mass (about \unit[15]{g}) in comparison to the wafer-like light absorbers (about $<\unit[1]{g}))$, the beakers have shown to be highly performing, achieving baseline resolutions $<\sigma_{BL}$ = \unit[9]{eV}~\cite{Reindl_beaker_2014}. Beakers offer a higher light collection efficiency in combination with the possibility to actively veto any surface-related alpha-induced backgrounds. 

Ge-wafers read out by NTDs instead find applications in the field of neutrinoless double-beta decay. The combination of Ge-NTD with Ge-absorbers turned out to be a good match providing baseline resolutions in $<{\mathcal{O}(\unit[100]{eV})}$~\cite{NTD_LDs_2015,0nubb_review_agostini_2023}.

\section{Benefits and Challenges of Low-temperature Detectors}
\label{bc}
 
Cryogenic detectors are very well suited for rare events search and allow us today to look back to about 35 years of successful employment in this field. However, realizing the full potential of these detectors is not free of challenges.

As outlined in Sec.~\ref{LTDs}, cryogenic detectors measure temperature changes at $\mathcal{O}(\mu\text{K}$)-level and low-temperature technology and vacuum techniques are a mandatory asset to provide the conditions for the milliKelvin temperature regime of detector operation. The apparatus which deliver such low temperatures, the so-called dilution refrigerators, are nowadays commercially available. In particular, the fast-growing field of quantum computing expanded the employment of this technology beyond fundamental particle physics and solid-state physics, resulting in reliable, partially automated and robust user-friendly systems.

The energy deposited by a particle in an absorber is, depending on the absorber material, shared between different excitations including e-h pairs, electrons, phonons, magnons, and others. A semiconductor detector can access only that part of the energy that is converted into e-h pairs, typically around 30\% of the energy deposition for electron recoils and strongly quenched values for nuclear recoils. A phonon detector instead has access to almost all excitations since the respective processes occur within the time scale of a typical phonon signal, that is in the regime of microseconds to milliseconds. This enhanced sensitivity constitutes a  key feature of this technology.

Not being restricted to semiconductors opens up the possibility of choosing an absorber consisting of multiple elements. Depending on the mass of the DM particle, a lighter/heavier target element can, due to the kinematics of the elastic-scattering process, improve the sensitivity significantly. Having a compound consisting of both, light and heavy elements, as e.g.~CaWO$_{4}$, enables improved sensitivity in a larger DM mass range. 

The most powerful feature of scintillating phonon detectors is the possibility for particle identification on an event-by-event basis. Considering that the dominant fraction of the background in low-temperature detectors comes from $\beta$/$\gamma$-particles which, at the same energy, produce significantly more scintillation light than the sought-for nuclear recoils, the simultaneous readout of phonon and light signal is a powerful tool for background rejection. The detection of the scintillation light requires sensitive cryogenic light absorbers and a good light collection efficiency. Since the amount of deposited energy going into light production is only at the ${\mathcal{O}(\text{1\%})}$, there is an ultimate limitation of the discrimination power going to lower energy. To make the best use of particle identification the energy deposit should be $>\unit[1]{keV}$. Thus looking for the classical WIMP was a great fit for these experiments. 
Employing absorbers with multi-elements allows us to observe a potential signal simultaneously on different nuclei since interactions on different nuclei can, to some extent, be discriminated against each other. This is particularly interesting in the case of a positive signal since it allows studying spectral shapes on different  elements and investigate characteristics of the possibly discovered new particle. As shown in Fig.~\ref{f:DMinCryoDet}, for the lowest DM masses, the larger energy transfer to the low-A targets results in a signal spectrum that extends further above the detector energy threshold. By reducing these thresholds, this constraint is relaxed and the detection can benefit from the large $A^2$-enhancement of the rate provided by the heaviest element.

The region of interest for nuclear recoil interactions vs. electron scattering can be set by in-situ measurements of the light quenching factors using a neutron source (see Fig.~\ref{f:PID_CRESST}). This is a great advantage of the low-temperature detectors exploiting phonon-light technique in comparison to single-channel detectors, e.g.~measuring only the light or charge signal from an interaction.

Low-temperature detectors are also very flexible and versatile. To-date the technology has successfully moved from classical WIMP search in the direction of light DM where nuclear recoil baseline resolutions $\sigma_{BL}$ of only a few ${\mathrm{eV_{nr}}}$ have been demonstrated (e.g. $\sigma_{BL}<\unit[5]{eV_{nr}}$ in~\cite{CRESST_SiDet,HVeV2021Performance,CRESSTsurf2017,CPD2021DM}). Such excellent performance allows an experiment to be sensitive to ${\mathcal{O}(\unit[100]{MeV})}$-scale DM particles through nuclear recoils.

A common and big challenge of massive low-temperature detectors employing the phonon-light technique is the reproducible production of both the scintillating target crystals and the sensitive TESs to read out the signals. In order to meet the requirements for rare event search, the internal radioactivity level originating from trace contamination of radioactive isotopes from e.g.~the natural decay chains have to be kept at ultra-low levels $ \ll \mathcal{O}(\unit[1]{mBq/kg})$. This can be achieved by a very careful selection of the materials, starting from the raw materials for the crystal growth, as well as following well-defined protocols and a strict scientific methodology for the growth process including crystal shaping and final surface treatment. Besides the radioactive aspect, also the material properties of the crystal are of concern. Perfect single-crystal quality and crystal lattice structures that do not show inner tension are key to ensuring excellent phonon propagation properties.

When it comes to TES production, thin-film technology in combination with several iterations of wet chemistry and photo-lithography processes are necessary in order to equip the crystals with these ultra-sensitive multi-layer sensors. The superconducting tungsten and aluminum thin films are typically produced via electron-beam evaporation, or in advanced sputtering systems, the thin gold structures are fabricated by sputtering exclusively. As the sensors are typically directly fabricated onto the massive crystals, only $\mathcal{O}(1)$ crystal is normally processed at a time, inducing susceptibility to systematics. 
The main factors characterizing a TES are its transition temperature $T_{c}$ and the transition widths. In the case of W-TES this means sensors should exhibit a $T_{c}$ around \unit[15]{mK} and a transition width of \unit[$<$1]{mK}. In order to collect exposures of $\mathcal{O}(\unit[100]{\text{kg-days}})$ (see Fig.~\ref{f:CRESSTproj}) larger arrays of $\mathcal{O}(\unit[100]{detectors})$ are mandatory to avoid measurement times of $>\unit[5]{years}$.  
The reproducibility requirement has a strong impact on both the quality of the crystal and the sensor.\footnote{NTDs, as used in neutrinoless double-beta decay searches, are very suited for large detector arrays as they demonstrated consistent energy resolution at the Q-value of the decay at the MeV-scale.} The overall achievable sensitivity or limit from combined data sets from different detectors can only be improved significantly if the detectors have similar sensitivities. The sensitivity depends on a combination of parameters. The energy threshold, the background levels, and, if required, the accuracy of the background model are the dominant factors that, if compatible among detectors, can lead to improved results.

To paint a complete and consistent picture, it is interesting to confront the intrinsic properties and benefits of liquid noble gas detector to low-temperature detectors. In a liquid noble gas detector two signals (e.g. using a dual-phase time projections chamber (TPC)) can be detected:  the primary scintillation light pulse produced in the liquid and a secondary scintillation light pulse produced via a gas stage that converts ionization to light. Liquid noble gas detectors, which nowadays arrive at ultra-low background levels and exposures of $\mathcal{O}(\text{\unit[1]{tonne-years}})$ are easier to scale up due to the fluid nature of the target material. Furthermore, also the radiopurity of a liquid substance can be enhanced by dedicated and continuous purification processes that are superior to material selection in solids.

In addition, using the self-shielding effect of large liquid volumes allows to suppress background from the experimental setup and facility. This is another advantage of the noble gas detectors over cryogenic detectors that, also if employed in large arrays which can profit from a shielding effect, still need to rely on a holding system for both mechanical and thermal connections.

This list of benefits explains the leadership role that direct DM experiments based on noble liquids hold today for $>\unit[1]{GeV/c^2}$.

However, when it comes to detector performance the cryogenic detectors outperform the liquid noble gas detectors thanks to their intrinsic low energy thresholds. This is due to inherent energy threshold limitations for the liquid noble gas detectors, as the amount of produced scintillation light by a particle interacting in the liquid xenon or argon is only in the ballpark of $\mathcal{O}(10)$ photons per keV. Furthermore, xenon and argon are both very heavy nuclei, which results in less momentum being transferred from DM in comparison to target materials offering lighter scattering partners.

However, new highly relevant and demanding challenges have emerged as a result of achieving excellent low thresholds. One of them is gaining a detailed understanding of the detector response to nuclear recoils at the eV scale~\cite{REVIEW_QFS_2023}. Currently, available low-energy calibration sources are based on electron recoils while the DM signature in many direct detection experiments is nuclear recoils. Moreover, at recoil energies below several hundred eV a significant fraction of energy might get stored in crystal lattice defects and thus the resulting energy spectrum might get distorted by this effect~\cite{CrystalDefectsDM2022}. A novel approach to a direct energy calibration of nuclear recoils based on thermal neutron capture was proposed by the CRAB collaboration in~\cite{CRAB2021idea}. For some isotopes, when a nucleus captures a thermal neutron, there is a possibility that the de-excitation process will go through a single MeV-gamma emission, causing a mono-energetic nuclear recoil. For instance, the anticipated nuclear recoil energy deposition for the $^{182}$W isotope in this process is \unit[112.5]{eV}. This mono-energetic nuclear recoil peak has recently been observed around the expected value in CaWO$_4$ detectors: in a dedicated measurement by the CRAB and NUCLEUS collaborations~\cite{CRAB2023} and in neutron calibration campaigns of the CRESST-III experiment~\cite{CRESST2023ncal}. This method and its extension making use of timing information~\cite{CRAB2023timing} enables precise detector response study for different target materials, such as Al$_2$O$_3$, Si, Ge, and CaWO$_4$, and in combination with sources based on electron recoils can improve understanding of the signal formation at low energies.

Another to-date big challenge with major impact and possibly, if not mitigated, serious consequences for the low-threshold cryogenic detectors is the observed unexpected rises in the measured energy spectra at energies below several hundred eV. Such rises have been observed in various target materials using different phonon sensor techniques in both underground and surface facilities~\cite{ExcessWorkshop2022}. These spectral features, exceeding signals from expected background sources, are commonly referred to as low-energy excesses (LEEs). The LEEs populate the regions of interest for low-threshold solid-state cryogenic detectors and thus currently pose the main challenge in enhancing sensitivity to light DM. As an example, Fig.~\ref{f:CRESSTproj} demonstrates the impact of the observed LEE on the DM limits established by the CRESST-III experiment in~\cite{detA}, with further discussion provided in Sec.~\ref{sss:cresst_nextsteps}. The significance and global nature of the LEEs have prompted the organization of a series of dedicated community-wide EXCESS workshops, providing a platform for thorough discussions on the observed LEEs~\cite{EXCESSworkshop21,EXCESSworkshop22,EXCESSworkshop22idm,EXCESSworkshop23taup}. 
    
Certainly, there are unique detector-specific sources contributing to the observed LEEs. An example is the recently revealed dominant contribution of luminescence from printed circuit boards used in detector holders to the LEE observed in the SuperCDMS-HVeV detectors~\cite{SuperCDMSHVeV2022excess}. However, there are indications of shared properties among the LEEs observed by crystal cryogenic phonon detectors. For example, the LEE event rate decays with time~\cite{queguiner2018phdEdelweiss, CRESST2022LEE,SPICE2022stress} and is strongly enhanced following thermal cycles to tens of Kelvin~\cite{queguiner2018phdEdelweiss,CRESST2022LEE}. During these thermal cycles, detectors are warmed up to temperatures of $\mathcal{O}(\unit[10]{K})$ and then cooled back down to their operating temperatures around $\mathcal{O}(\unit[10]{mK})$, where data-taking can resume. To-date, a radiogenic origin of the LEEs has not been identified and can largely be excluded as a major contribution, given the observed increases in event rates due to thermal manipulations alone. Furthermore, the measurements of the EDELWEISS collaboration suggest that the LEE has a non-ionizing nature~\cite{edelweiss2022NbSiDM, EdelweissIII2016}. Therefore, the CRYOSEL project is proposing a novel way to tag athermal phonons emitted via the NTL process and as a result effectively reject non-ionizing events~\cite{cryosel_2023}. 

The community is actively investigating phenomena related to solid-state physics, such as relaxation processes due to stress induced by detector elements, thermal expansion or intrinsic crystal processes, as plausible explanations for the LEEs and their observed patterns. Stress may potentially be generated at the interfaces between an absorber and a sensor, at glued or bonded elements, or by the mechanical force of supporting structures. 

The SPICE/HeRALD collaboration has recently demonstrated that a crystal operated under higher stress conditions induced by a glue-based holding system, measures a much higher LEE rate below \unit[40]{eV} in comparison to a crystal suspended by bond wires, which minimizes stress from supporting structures~\cite{SPICE2022stress}. Therefore, reduction of the mechanical stress on the crystal is one of the LEE mitigation strategies. 

Another approach is to distinguish events potentially induced by stresses from the holding structures and recoils in the absorber volume. The CRESST~\cite{pucci_ltd2023} and NUCLEUS~\cite{nucleus_rothe_2020} experiments propose different versions of active holding structures instrumented with a TES. The simultaneous readout of the signal from the holders and the absorber is expected to allow the identification of events originating at the interfaces between the holders and the crystal. On the other hand, the segmentation of the BULLKID monolithic detector arrays should allow for the rejection of events originating outside each crystal dice~\cite{bullkid_lowenergy_2023}.

In order to distinguish particle interactions in the absorber from events related to the sensors themselves, e.g.~stresses associated with thin film fabrication processes or thermal expansion mismatch between the crystal and the sensor materials, signal readout from multiple phonon sensors attached to the same crystal is currently being explored. Preliminary results from several independent measurements have identified non-coincident LEE-like events associated with a single TES channel~\cite{2TESSpice,2tes_pucci_workshop,2TES_pucci}. While understanding the exact physical mechanism causing such events requires further investigation, identifying and excluding them from the measured energy spectra enables a partial suppression of the LEE.

Sub-eV energy resolution is required for the hunt of DM-electron scatterings. DM-electron interactions in scintillating targets are expected to stimulate the emission of single photons.  In this light, a first study of the interplay between particle and condensed matter physics was presented in \cite{zema202DM_electron} with the aim of defining the theoretical framework of this novel detection technology, which complements the existing searches based on CCD detectors.

\section{{Current Dark Matter Search Experiments Employing Scintillating Crystals at milliKelvin Temperatures}}
\subsection{CRESST}
\label{s:cresst}

\begin{figure}
        \centering 
        \includegraphics[width=0.8\columnwidth]{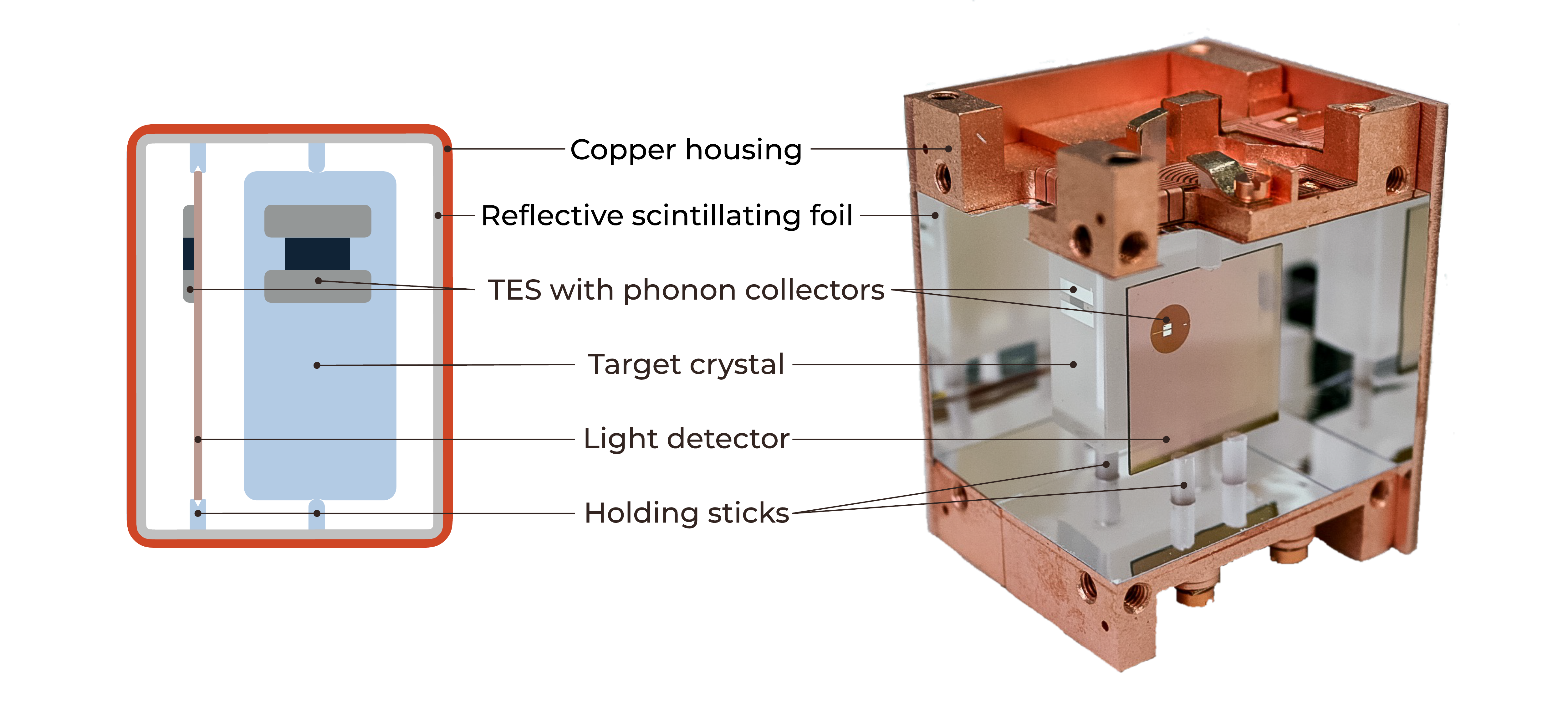}
\caption{Standard CRESST-III detector module featuring labeled main components. Left: schematic drawing, not to scale. Right: photograph of the module with two housing walls removed for better visibility, provided by the CRESST collaboration. The target crystal and wafer-shaped light detector are both equipped with a separate W-TES and mounted in a high-radiopurity copper housing. Reflective scintillating foil covers the housing walls to maximize light collection efficiency.}
\label{f:CRESSTScheme}
\end{figure}

The CRESST (Cryogenic Rare Event Search with Superconducting Thermometers) experiment combines the high sensitivity of cryogenic TES-based detectors with particle identification through scintillation light. Its first stage, CRESST-I (1996-2001), was among the pioneers of direct DM searches, employed a single-channel readout of the phonon signal from $\sim \unit[250]{g}$ sapphire target crystals \cite{angloher_limits_2002}. In the second stage, CRESST-II (2002-2015) enabled background reduction in the keV-energy range by taking advantage of using scintillating CaWO$_4$ crystals as targets. A two-channel readout system was implemented to register both the phonon and light signals after a particle interaction within the $\sim \unit[300]{g}$ absorber \cite{cresst2012results}. Additionally, the presence of the heavy tungsten nucleus was an advantage for $\mathcal{O}(\unit[10]{GeV/c^2})$ WIMP searches due to the $A^2$-enhancement of the scattering cross-section (Eq.~\ref{eq:dSigmadE}). The main focus of the ongoing, third phase of the experiment, CRESST-III (from 2016), is sub-GeV DM search, for which thresholds of $\mathcal{O}(\unit[10]{eV})$ are required. To achieve these low thresholds, the dimensions of the target crystals in CRESST-III were reduced following the arguments from Sec.~\ref{s:sensors}, with standard modules now operating with \unit[24]{g} CaWO$_4$ crystals. For sub-GeV DM, the naturally present light oxygen nucleus becomes the main probe due to the higher expected recoil energy reach (see Fig.~\ref{f:Rate}). While particle discrimination based on scintillation light is no longer effective below $\sim \unit[1]{keV}$, the light channel offers precise in-situ background measurements and identification of different background components at higher energies, crucial as input for background simulations. Furthermore, the wafer-like light absorbers, when considered as main targets, enable even lower energy thresholds.

The standard CRESST-III detector module, shown in~Fig.~\ref{f:CRESSTScheme}, employs a \unit[($20\times20\times10$)]{mm$^3$} CaWO$_4$ crystal as the target.
Scintillation light, with maximum of the emitted spectra at ${\sim\unit[420]{nm}}$ for CaWO$_4$~\cite{CaWO4_scint_spectrum_2004}, is measured by a separate wafer-shaped silicon-on-sapphire \unit[($20\times20\times0.4$)]{mm$^3$} light detector that faces the target crystal. 

The target crystal and the light detector are installed in an encapsulated housing made of high-radiopurity copper. The walls of the module are covered with reflective and scintillating Vikuiti\textsuperscript{\texttrademark} foil to enhance the light collection. Both crystals are supported by a set of three CaWO$_4$ sticks, resulting in a fully scintillating inner housing. 

The target crystal and the light detector are each equipped with a W-TES with aluminum phonon collectors for signal read out~\cite{strauss_2017_CRESSTproto}, which are schematically shown in Fig.~\ref{f:PC} and discussed in Sec.~\ref{s:sensors}. 
The tungsten and aluminum structures are electron-beam evaporated onto the crystal surfaces. Detectors are operated at a temperature of approximately $\sim\unit[15]{mK}$, which corresponds to the transition temperature of tungsten in its $\alpha$ crystalline phase. The W-TES is weakly coupled to the heat bath via a gold stripe sputtered onto the crystal. The resistance change resulting from the temperature rise after energy deposition in the crystal is measured by the SQUID-based read-out circuit shown in Fig.\ref{f:SQUIDReadOut}~\cite{CRESST2009Commissioning}. Additionally, each detector is equipped with a gold ohmic heater to inject artificial pulses of fixed discrete energies. Response of the sensors to these pulses serves as input for the real-time detector stabilization algorithm, which keeps the detectors at their optimal operating temperatures, and for precise, time-sensitive determination of the detector response function over the full dynamic range~\cite{cresst2012results}. 

The CRESST experiment is running in a $^3$He/$^4$He dilution refrigerator in a deep-underground facility of the Laboratori Nazionali del Gran Sasso (LNGS), where excellent shielding against cosmic radiation is provided. The detector modules are protected from neutrons, gamma backgrounds and residual cosmic muons by several concentric layers of active and passive shielding materials~\cite{cresst2012results}. The current configuration of the CRESST setup allows the simultaneous operation of about 15 standard detector modules.

The Geant4-based physics simulation code, ImpCRESST, has been developed within CRESST to describe the expected electromagnetic backgrounds~\cite{CRESST2019bckmodel,CRESST2019bckmodelErr}. A recently refined normalisation method for the background model gained an improvement of 18.6\% in comparison to the previously used method~\cite{CRESST2023bckmodel}; the background model can now explain 82.7\% of the experimental backgrounds in the energy range between 1 and \unit[40]{keV}. While publicly available simulation packages are not validated with experimental data for materials used in CRESST at the \unit[10]{eV}-energy scale, the recent ELOISE project aims to improve the reliability of simulations for electromagnetic interactions at low energies in CaWO$_4$ and Al$_2$O$_3$~\cite{kluck_eloise_2023}. 

Background descriptions from the simulations can be incorporated into the framework for obtaining DM results based on a profile likelihood ratio test recently presented in~\cite{cresst_likelihood_2024}. However, the DM exclusion limits reported by CRESST so far and mentioned in this review are conservatively calculated using the Yellin method~\cite{yellin, yellin2}, which treats all events in the acceptance region as a potential signal.

The intrinsic radiopurity of the crystal significantly contributes to the total background level~\cite{CRESST2023bckmodel}. Moreover, the properties of the target crystal influence the amount of available scintillating light required for particle identification~\cite{strauss2014_qf}. The CaWO$_4$ crystals used in CRESST are grown in-house at the Technical University of Munich~\cite{Erb2013}. By maintaining control over each production step, it was possible to achieve high radiopurity, good optical quality, and minimal crystal lattice stress compared to commercially produced crystals~\cite{Strauss_2015, muensterPhD}. Furthermore, extensive chemical purification of the raw materials was performed to grow the CaWO$_4$ target crystals used in the most recent data-taking campaign described in Sec.~\ref{s:cresst_nextcampaign}. This resulted in a further reduction of the $\alpha$-decay rate from the natural decay chain by at least a factor of 6 compared to the crystals grown without this additional purification step~\cite{cresst2023cawo4bck}. With this advancement, the crystal's intrinsic radioactive background became a subdominant component of the overall background rate. 

\begin{figure}
        \centering 
        \includegraphics[width=1\columnwidth]{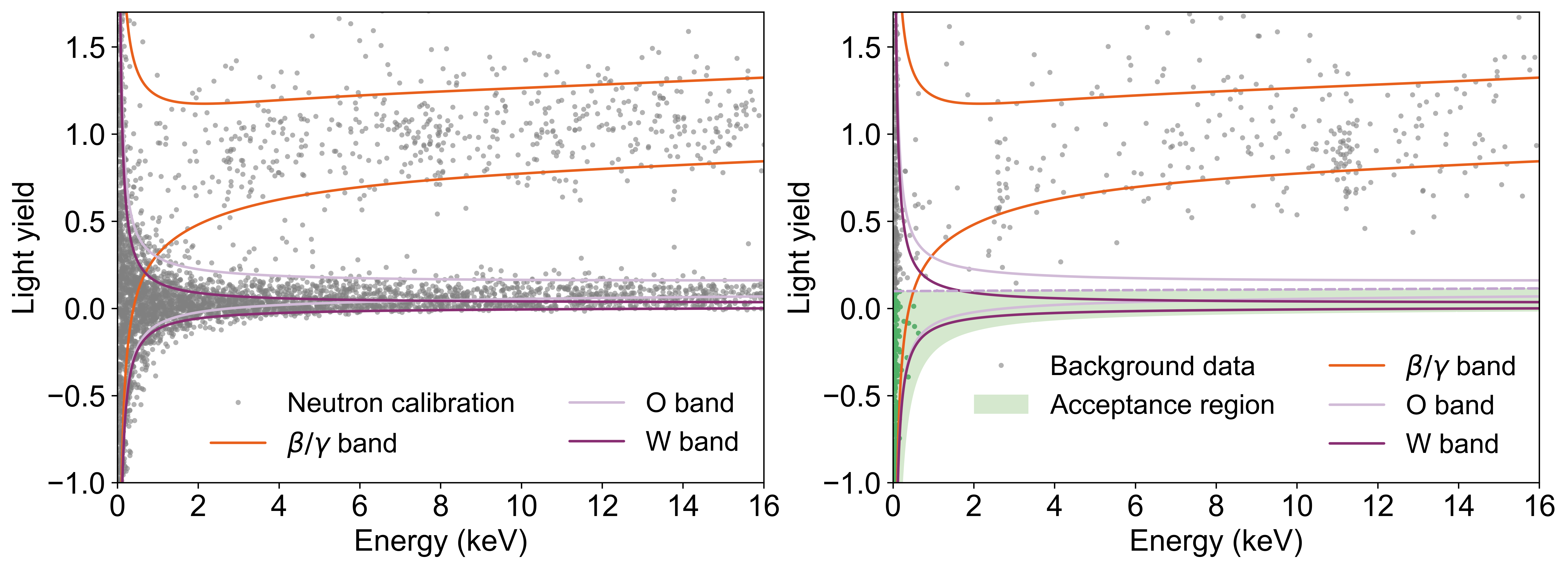}
\caption{Light yield versus energy for the CaWO$_4$ detector in the first CRESST-III data-taking campaign. Left: Neutron calibration data are shown as gray dots. These data were fitted to obtain the description of the recoil bands following~\cite{strauss2014_qf}: orange for $\beta/\gamma$ events, light purple for oxygen recoils, and purple for tungsten recoils. The calcium band is omitted for clarity. The lines indicate the 90\% upper and lower boundaries of each band. Right: Background data used for the DM search are shown as gray dots. The solid lines are identical to the ones in the left panel. The green shaded area highlights the acceptance region defined from the mean of the oxygen band (light purple dashed line) down to the 99.5\% lower boundary of the tungsten band. Green dots denote events within the acceptance region. Figure adapted from~\cite{detA}.}
\label{f:CRESST_LY}
\end{figure}

\subsubsection{First CRESST-III Data-Taking Campaign: Achieving a \unit[30]{eV} Energy Threshold extends Sensitivity to sub-GeV DM}
In the first CRESST-III data-taking campaign (2016-2018), a groundbreaking energy threshold of \unit[30.1]{eV} was achieved with one of the operated \unit[24]{g} CaWO$_4$ detectors~\cite{detA}. The light channel enabled particle identification. The left panel of Fig.~\ref{f:CRESST_LY} shows the $LY$ versus energy of events during the neutron calibration campaign with an AmBe source, conducted to precisely characterize the $LY$ over a wide energy range. Two clearly distinguishable bands are observed: electron recoils originating from $\beta/\gamma$-events with $LY\approx1$ and nuclear recoils from neutron interactions with $LY$ close to 0. Since the amount of scintillating light depends on the mass of the target nucleus, the nuclear recoil band for CaWO$_4$ contains three components. The bands are fitted following the description in~\cite{strauss2014_qf}, and 90\% boundaries are depicted as solid lines in Fig.\ref{f:CRESST_LY}. The fraction of deposited energy converted into light production for $\beta/\gamma$-events in such CaWO$_4$ detectors was determined to be ${(6.6\pm0.4)\%}$\cite{CRESSTII2014}. 

\begin{figure}
        \centering 
        \includegraphics[width=0.58\columnwidth]{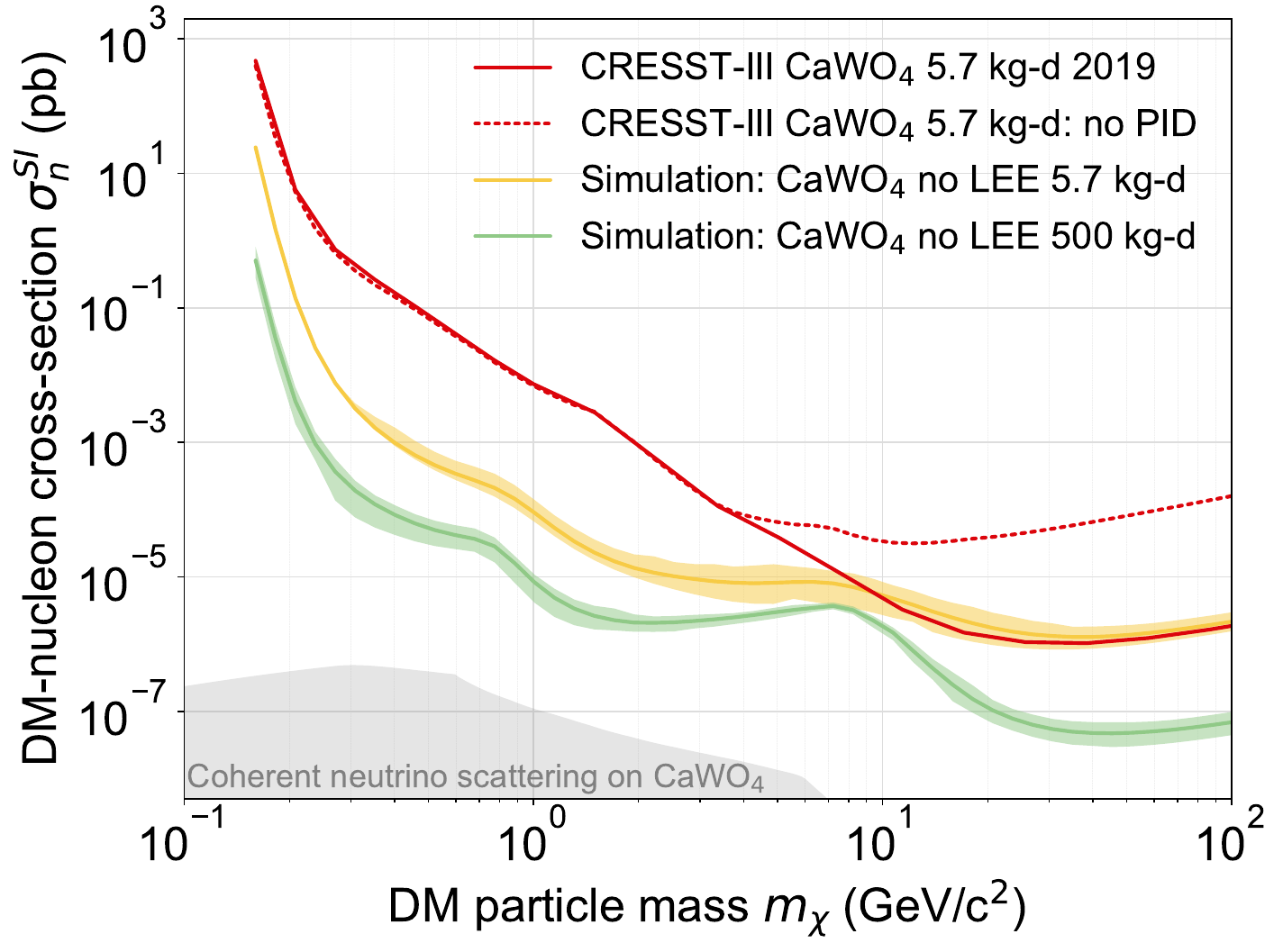}
\caption{
Expected sensitivity of the CRESST-III \unit[24]{g} CaWO$_4$ detector with a \unit[30]{eV} threshold to SI elastic DM-nucleus scattering across various DM masses. The red line represents the upper limit on the interaction cross-section at a 90\% confidence level, derived from data collected during the first CRESST-III data-taking campaign reported in~\cite{detA}. The red dashed line shows the limit calculated with the same dataset without considering information from the light channel by accepting all events with $LY\in(-10,10)$ to simulate a phonon-only read-out experiment with particle identification disabled. The yellow line illustrates the limit obtained with a simulated spectrum that follows the measured one but is modified to exhibit a flat background level down to the threshold, showing the influence of the LEE on the sensitivity. The green line shows the limit derived from a simulated spectrum excluding the LEE and featuring an increased gross exposure of \unit[500]{kg-days}. Shaded areas indicate the 90\% confidence interval borders for the corresponding limits. For the simulated limits, the same detector performance as in~\cite{detA} was assumed.}
\label{f:CRESSTproj}
\end{figure}

The right panel of Fig.~\ref{f:CRESST_LY} shows the $LY$ versus energy of the background data used for the DM search. The acceptance region is shaded with green. Clear separation of the bands for electron recoils and nuclear recoils provided a powerful tool for background suppression above $\sim\unit[1]{keV}$, leading to zero registered events in the acceptance region above \unit[630]{eV}. With this detector, CRESST-III explored new DM parameter space for DM masses down to $\unit[0.16]{GeV/c^2}$, depicted as the red solid line in Fig.~\ref{f:limits} and~\ref{f:CRESSTproj}~\cite{detA}. To date, these limits remain the most stringent set via the SI elastic scattering of DM with the target material’s nuclei for sub-GeV DM down to $\unit[0.165]{GeV/c^2}$.

The impact of the particle identification on the resulting limits is shown with the dashed red line in Fig.~\ref{f:CRESSTproj}. To imitate the absence of the light channel, all registered events with $\text{LY}\in(-10,10)$ that survived the selection criteria were considered for the limit calculation using the Yellin method~\cite{yellin, yellin2}, instead of restricting to events in the acceptance region. For $m_\chi>\unit[5]{GeV/c^2}$ particle interactions in the absorber deposit energies of $\mathcal{O}(\unit[1]{keV})$ and thus event discrimination becomes possible. Therefore, the background level gets suppressed leading to an improvement of the exclusion limit by up to two orders of magnitude when the light read-out channel is enabled.

In this first data-taking campaign of CRESST-III, together with achieving detection thresholds of ${\mathcal{O}(\unit[10]{eV})}$, an unforeseen rise of the energy spectra was observed below \unit[200]{eV} - the LEE introduced in Sec.~\ref{bc}. The LEE significantly populates the region of interest for DM particles with ${m_{\chi}<\unit[10]{GeV/c^2}}$. Therefore, identifying the origin of the LEE and mitigating its impact are currently the primary goals of the CRESST experiment. 

\subsubsection{Next CRESST-III Data-Taking Campaign: A Multi-Target Approach for LEE Studies and New DM Results}\label{s:cresst_nextcampaign}

The CRESST-III data-taking campaign, started in the middle of 2020 and concluded in early 2024, was specifically dedicated to investigating the LEE. The flexibility of the CRESST technique allows to operate different crystal materials as targets. In addition to  CaWO$_4$, CRESST installed Al$_2$O$_3$, LiAlO$_2$, and Si detectors to examine the material dependence of the LEE. Some additional modifications to holding structures and module's housing were made. See Ref.~\cite{CRESST2022LEE} for an overview.

With the help of those modules several new aspects of the LEE were revealed~\cite{CRESST2022LEE}. First of all, the LEE is present in detectors with all the different target materials used. Secondly, the LEE rates do not scale with the mass of the target crystal. Thirdly, the LEE event rate decays exponentially with time since the cool-down. Moreover, thermal cycles to tens of Kelvin enhance the LEE rate by up to an order of magnitude. Based on these observations, external and intrinsic radioactivity, as well as DM, were excluded as major origins of the LEE. The favored hypothesis for the LEE origin is currently stress-related processes induced by the holding structures and/or the sensors' films.

Despite the presence of the LEE, CRESST-III obtained several new results on DM in this data-taking campaign with different target materials. The LiAlO$_2$ target contains three isotopes sensitive to SD DM interactions: $^6$Li, $^7$Li, and $^{27}$Al. Using the LiAlO$_2$ crystal as an absorber, CRESST-III set the currently strongest limits on SD interaction of DM particles with protons and neutrons for the mass region between 0.25 and ${\unit[1.5]{GeV/c^2}}$~\cite{CRESST2022Li}. 

The wafer detectors, normally used as ancillary channels for light collection, can also be used as main targets. Due to their small sizes, they achieve even lower detection thresholds and thus push the sensitivity to lower DM masses. A silicon wafer detector achieved the energy threshold of \unit[10.0]{eV}. With the data from this miniature detector, CRESST-III extended the reach to lower DM masses and improved the limits between 130 and ${\unit[160]{MeV/c^2}}$ by a factor of up to 20 compared to previous best results in this mass range~\cite{CRESST_SiDet}. The limit is shown with solid orange line in Fig.~\ref{f:limits}. A silicon-on-sapphire wafer detector achieved an even lower energy threshold $<\unit[10]{eV}$ which enables probing DM masses below $\unit[100]{MeV/c^2}$~\cite{cresst_singlephotons_2024}.

\subsubsection{Future Directions: New Detector Concepts, Target Materials, and Exposure Increase}
\label{sss:cresst_nextsteps}

Based on the gained knowledge about the LEE, CRESST developed several new detector module concepts for future data-taking campaigns~\cite{pucci_ltd2023}. The modifications mainly fall into two categories. The first group focuses on modifications to the support structures. By reducing mechanical stress on the crystal its influence on the LEE rate can be investigated. If the holding structures are instrumented with TES themselves, it should be possible to discriminate events originating at the contact points. The second type of modifications exploits discrimination of events in the absorber from events in the TES films or material interfaces by reading out the signal from the same crystal by two separate TES sensors simultaneously. First above-ground measurements with such doubleTES detectors show promising results in identifying a contribution of TES-related events in the measured low energy spectrum~\cite{2TES_pucci}. In addition to the LEE-oriented modifications, new detector designs target an enhanced DM sensitivity by further improving the thresholds and implementing a 4$\pi$ active veto of the target crystal. The detectors featuring these new designs will be operated in the forthcoming CRESST-III campaign, scheduled for 2024.

If the LEE's effects are mitigated, a strong sensitivity boost for ${m_{\chi}<\unit[10]{GeV/c^2}}$ is expected. To illustrate this effect, we use a description of the measured energy spectrum that led to the 2019 SI DM-nucleon interaction limits~\cite{detA} using the likelihood framework for background modeling~\cite{cresst_likelihood_2024}. In this model, the LEE is approximated by an exponential function. We then simulate the energy spectrum according to this analytical description, but without the LEE component. Thus the simulated spectrum represents the situation where the measured spectrum stays roughly flat down to the threshold. The DM exclusion limit calculated at the 90\% confidence level with the Yellin method~\cite{yellin,yellin2} for this simulated data, assuming the same detector performance and exposure as in~\cite{detA}, is shown with the yellow line in Fig.~\ref{f:CRESSTproj}. While for ${m_{\chi}>\unit[10]{GeV/c^2}}$ the limit is not influenced by the absence of the LEE, for the lighter DM the sensitivity is improved by up to two orders of magnitude.

Additionally, new target materials are being considered within CRESST. To improve sensitivity to SD DM interactions with CaWO$_4$ targets, CRESST is currently exploring the possibility of $^{17}$O enrichment of the crystals~\cite{kinast2023enrich}. A new attractive material candidate is diamond. Due to its high Debye temperature it has superior phonon propagation properties. A diamond crystal operated as a cryogenic detector at a surface facility achieved the energy threshold of \unit[16.8]{eV} and showed a high potential for sub-GeV DM search~\cite{diamond2022,diamond2023DM}. 

Meanwhile, a major upgrade of the CRESST setup at LNGS in order to increase the number of available read-out channels is in preparation. While new SQUIDs and wiring are already procured, the new DAQ and electronic modules are in the pre-production phase. The planned outcome of this upgrade is a total of 288 read-out channels. This significantly increases the number of detectors that can be operated simultaneously and thus the achievable exposure. 

We estimate the expected DM sensitivity of a CaWO$_4$ detector in case of increased exposure and no LEE observed in the energy spectrum. The 90\% confidence level DM exclusion limit calculated with the Yellin method~\cite{yellin,yellin2} for the same simulated energy spectrum as in the `no LEE' case assuming \unit[500]{kg-days} of gross exposure is shown in Fig.~\ref{f:CRESSTproj}. A sensitivity enhancement by more than an order of magnitude over a wide range of DM masses is expected. 

Although particle identification based on scintillation light is crucial for the CRESST sensitivity to DM masses above $\unit[1]{GeV/c^2}$, its capacity is naturally limited at low energies, e.g. only a single photon is expected to be generated in CaWO$_4$ by a $\beta/\gamma$-interaction of \unit[50]{eV} energy. At the same time, further reduction of the threshold will allow probing DM masses below $\unit[100]{MeV/c^2}$. The exact detector design for the phase following the upgrade is yet to be determined, as it heavily relies on the outcomes of the upcoming data-taking campaign and the progress made in understanding the LEE.

\subsection{COSINUS}
\label{s:cosinus}

\subsubsection{The Physics Case}
The key challenges faced by direct DM detection experiments as e.g.~CRESST are a small anticipated energy transfer of $< \mathcal{O}$(\unit[10]{keV}), an almost featureless exponentially falling recoil energy spectrum and, a generally very low event rate (see Fig.~\ref{f:Rate}). 

Besides searching for a DM signal above common detector backgrounds, DM can also be identified via an annual modulation signal caused by the seasonal variation of the Earth’s velocity with respect to the sun and, thus, the DM halo~\cite{Freese_2013}. Aiming to exploit experimental techniques suitable for detecting a modulation of the event rate seems attractive as the correct phase and period are hard to be mimicked by backgrounds and, consequently, a robust handle for background rejection. On the other hand, the small change expected in the event rate per annual cycle requires experiments with a large target mass and a very low intrinsic background compared to pure signal-above-background-searches, which commonly use particle discrimination to search for the DM-nucleus interactions. 

A pioneer using the modulation as DM signature is the DAMA/LIBRA experiment which operates \unit[250]{kg} of Tl-doped sodium iodide (NaI(Tl)) crystals as room-temperature scintillators~\cite{bernabei_final_2013,bernabei_first_2018}. Over more than 25 annual cycles, their data confirm a modulating event rate, with a  statistical significance of \unit[13.7]{$\sigma$} that is compatible in phase and period with the expectation for a halo of DM particles in our galaxy~\cite{bernabei2022recent}. 

However, the DAMA/LIBRA result is in great tension with the other searches, as seen in Fig.~\ref{f:limits}, when using simple standard assumptions for a comparison \cite{lewin_smith_1996}, in specific for SI interactions and a Maxwell-Boltzmann velocity distribution for DM.

To avoid making assumptions on the particle physics properties of DM, the best way is to do an experiment with the same target crystal, NaI, and to provide an independent test of the DAMA/LIBRA result. As of today, such a material and model-independent unambiguous cross-check is still absent.

Currently, multiple experimental efforts are ongoing ~\cite{ANAIS_3y_2021,COSINE_3y_2022} or under construction~\cite{angloher2021simulationbased, Fushimi:2021lrd, Antonello:2020xhj,barberio2022simulation}. The Cryogenic Observatory for SIgnals seen in Next-generation Underground Searches (COSINUS) is one of them, albeit the only one operating NaI crystals as low-temperature scintillating calorimeters at milliKelvin temperatures and read out using TESs. Thanks to the phonon-light technique COSINUS is the only NaI-search with particle identification on an event-by-event basis. Furthermore light quenching factors can be determined in-situ and allow to discriminate between Na and I recoils due to their different $LY$.

\subsubsection{First Results from the remoTES Design}

Sodium iodide (NaI) crystals are very hygroscopic and thus do not allow for the fabrication of the TES directly onto the surface as done in CRESST-III~\cite{detA} and SuperCDMS~\cite{CPD2021Performance}. In these experiments, the absorber has to withstand the conditions of thin-film production via electron-beam evaporation and/or sputtering (high temperatures  and multiple thermal cycles, plasma, etc), wet-chemistry (etching) and several iterations of photo-lithography (solvents, water, exposure to air, UV light) to produce a multi-layer thin film that constitutes the TES. This restricts the application of TES to only robust standard materials.  To overcome this limitation a new TES-based detector concept, denoted remoTES was developed. Initially theoretically proposed in~\cite{pyle2015remoTES}, the first experimental proof of this particular readout design was recently demonstrated in~\cite{COSINUS2023FirstRemoTES}.

\begin{figure}
        \centering \subfigure[]{
        \includegraphics[width=0.4\columnwidth]{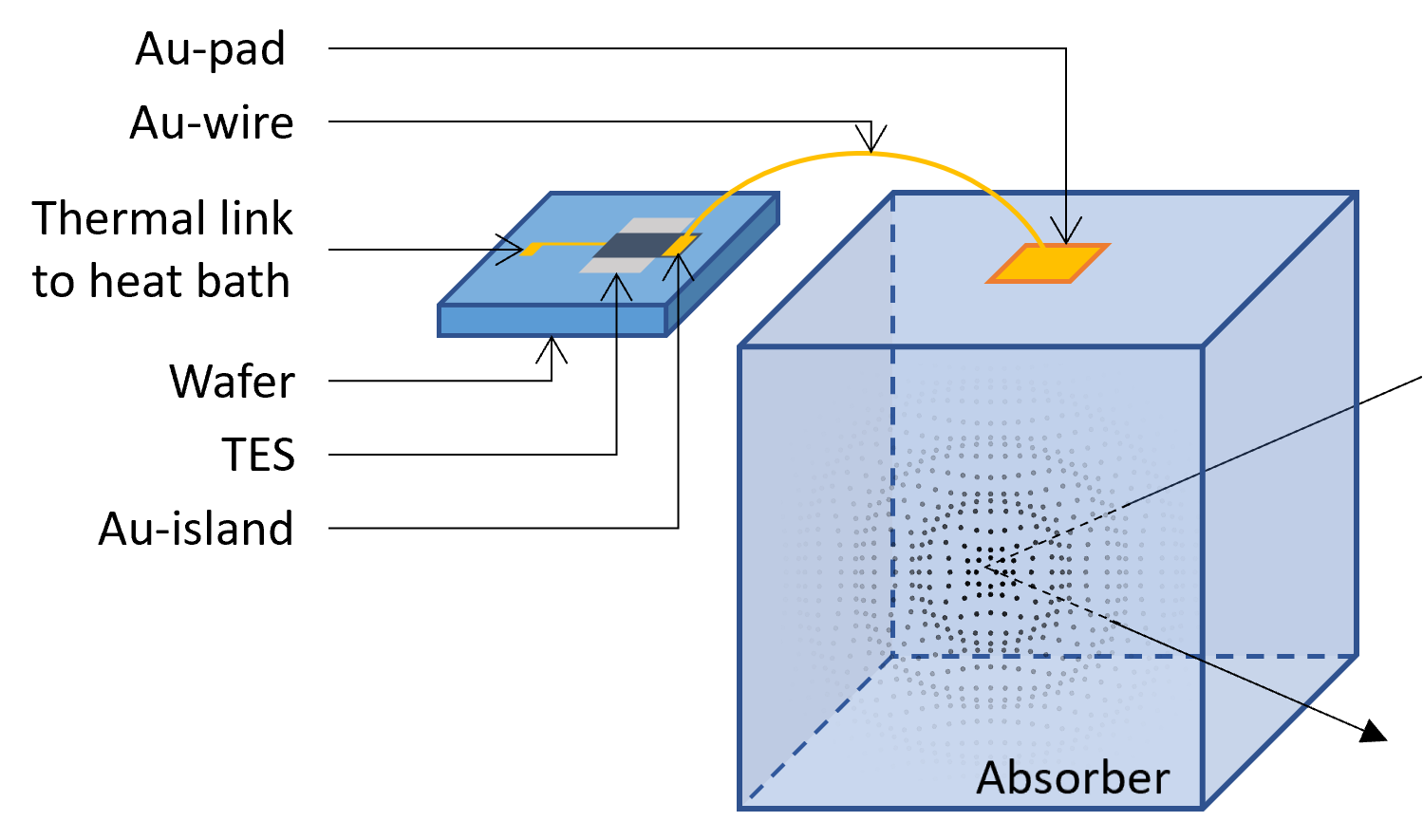}\label{f:remoTES}}
        \subfigure[]{
        \includegraphics[width=0.45\columnwidth]{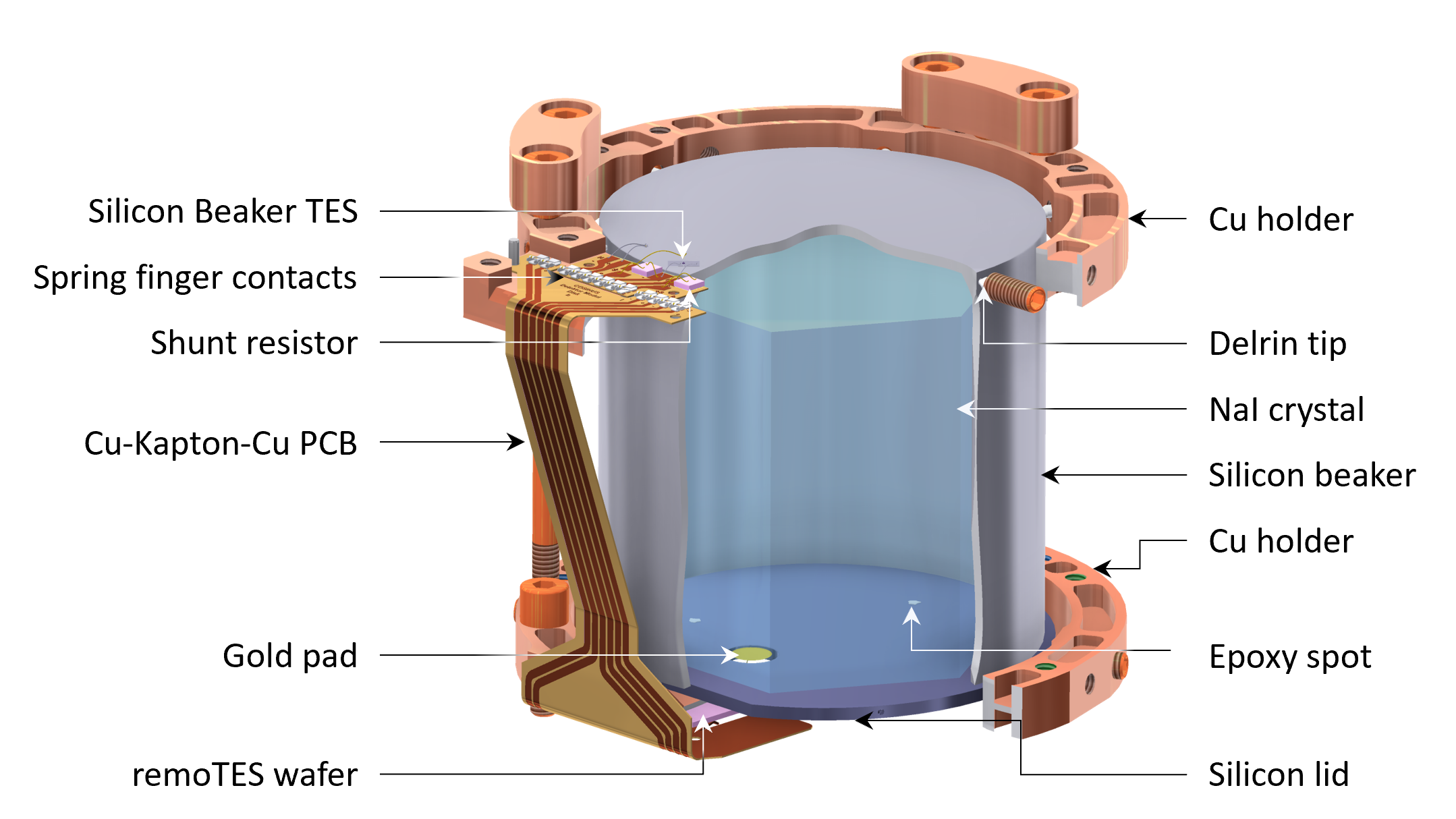}\label{f:module}}
\caption{(a) Depiction of the remoTES layout that allows to read out delicate absorber crystals like NaI. (b) CAD view of a NaI detector module consisting of a hexagonal NaI crystal of $\sim$\unit[100]{g}, read out by the remoTES layout. For efficient detection of scintillation light, the crystal is surrounded by a Si beaker, also read out by a W-TES.}
\label{f:COSINUS_concept}
\end{figure}

As depicted in Fig.~\ref{f:remoTES}, in a remoTES, the W-TES is fabricated on a wafer that is separated  from the absorber crystal. A gold pad located on the absorber crystal constitutes the phonon collector and is coupled to the TES via a gold wire-bond. 
This "remote" coupling scheme has two distinct advantages: firstly, the sensor and the  NaI absorber are physically separated and thus the NaI can be excluded from the fabrication process of the TES. Secondly, the phonons generated in the absorber couple directly to the electronic system of the gold pad via electron-phonon coupling  - that is particularly strong in Au - and also signal loss due to e.g.~acoustic mismatch, as experience in the composite design,~\cite{CRESST2009Composite} is reduced. The transmission probability for phonons from the NaI absorber via a glued sapphire wafer, equipped with a TES, onto its surface is only ${\mathcal{O}(\text{1\%})}$.

A first underground measurement of a NaI scintillating calorimeter was carried out mid 2022 in a test facility at the Gran Sasso underground laboratory. The phonon detector consisted of a \unit[1]{cm$^3$} radiopure NaI crystals (\unit[3.67]{g},  \unit[$\sim$ 10]{ppb} for $^{40}$K, SICCAS).
The Au-pad was made by a circular \unit[1]{$\mu$m} thick Au-foil (radius of \unit[0.75]{mm}) and was glued onto the crystals' surface using an epoxy resin.  A \unit[17]{$\mu$m} thick gold bonding wire was used to couple the gold pad to the thermometer which is a W-TES fabricated onto an Al$_2$O$_3$ wafer with dimensions \unit[(10x20x1)]{mm$^3$}.
To collect the scintillation light a beaker-shaped light absorber made from optically polished silicon with a mass of \unit[15.38]{g} was employed to encapsulate the NaI. At W-TES deposited directly on the Si beaker was used to register the light signal. 

The NaI phonon detector showed a baseline resolution of ${\sigma_{BL} = \unit[0.441 \pm 0.011]{keV}}$ which would allow triggering below \unit[2]{keV} in a low-background environment with the typical relation of $E_\text{thr} \approx 5 \sigma_{BL}$. However, to minimize pile-up, an analysis threshold of \unit[4]{keV} was set~\cite{cosinus_collaboration_deep-underground_2023}.

\begin{figure}[t]
\centering
\includegraphics[width=1\textwidth]{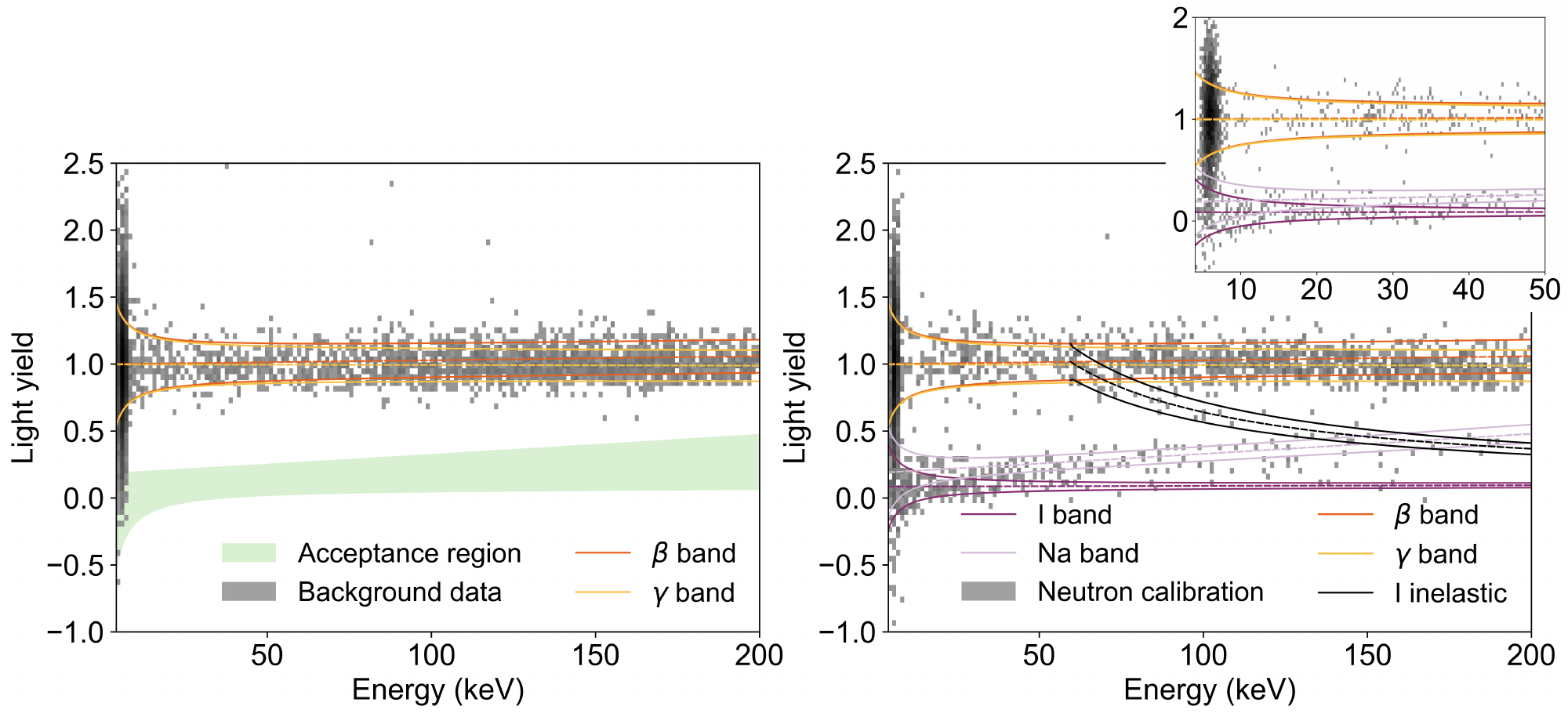}
\caption{2D Histogram of light yield vs.~the total deposited energy for both background (left) and neutron calibration data (right). Both figures show the fit to the $\beta$ (orange) and the $\gamma$ band (yellow) as found by a combined maximum likelihood fit. Together with the neutron calibration data the fit results for the nuclear and inelastic recoil bands are shown. The right panel includes a zoom into the low energy region to visualize the particle identification and the possibility of discriminating Na recoils from I recoils to a certain extent. The densely populated region around \unit[6]{keV} in the $\beta$/$\gamma$ band corresponds to the $^{55}$Mn K$_{\alpha}$ and K$_{\beta}$ lines used for energy calibration. The light green shaded region in the left panel marks the acceptance region for DM inference with the Yellin method. Figure adapted from~\cite{cosinus_collaboration_deep-underground_2023}.}
\label{f:lightyield_COSINUS}
\end{figure}

In the left plot of Fig.~\ref{f:lightyield_COSINUS} background data are shown in the $LY$ vs.~energy plane. At $LY\approx1$, the event distribution due to $\beta/\gamma$-interactions in the NaI is visible. The dense region at around \unit[6]{keV} is due to X-rays shining on the NaI absorber from a $^{55}$Fe source. 
On the right, instead, data from a calibration with an AmBe-neutron source are depicted. In this 2D histogram an additional band due to nuclear recoil events at lower $LY$ values appears in comparison to the left plot. A true separation of  $\beta/\gamma$-events from the quenched nuclear recoil events is demonstrated.  The events visible above \unit[50]{keV} between the two event distributions can be attributed to inelastic scatterings off iodine nuclei. This measurement is the first demonstration of event-by-event discrimination in a NaI-based detector \cite{cosinus_collaboration_deep-underground_2023}. 

In addition, the full energy-dependent quenching behavior measured in-situ and down to the (analysis) threshold can be extracted from the fit in Fig.~\ref{f:lightyield_COSINUS}, a great advantage of the phonon-light technique in comparison to the DAMA-like experiments. The QF at \unit[10]{keV} gives $QF_{\text{Na}}(10\,\text{keV}) =0.2002\pm 0.0093$ and $QF_{\text{I}}(10\,\text{keV})=0.0825\pm 0.0034$, allowing to distinguish even between recoils off sodium and iodine to a certain extent. An excellent light detector with an baseline resolution in the order of \unit[5]{eV} is crucial to allow to distinguish nuclear recoil signal events from common background down to an energy threshold of \unit[1]{keV}.

It should be mentioned, that to compare different DAMA-like searches, a precise knowledge of the detector response to nuclear recoils is key. To date, the quenching factors observed for nuclear recoil events in Tl-doped crystals operated as room temperature scintillators disagree between different studies~\cite{REVIEW_QFS_2023}. Both the influence of the Tl-dopant level and the functional dependence of the light response due to intrinsic properties of the crystals are not settled and only recently systematic studies have been undertaken. For DAMA-like searches, this uncertainty on the light response directly translates to the inferred nuclear recoil energy scale. A direct comparison among scintillation-light-only experiments, therefore, still suffers from significant systematics.

The final mass of the COSINUS target crystals is not yet fixed and will be guided by the achieved performance (energy threshold) of the NaI calorimeter. For \unit[100]{g} crystals \unit[1000]{kg-days} would mean a measurement time of about two to three years with 20 working detector modules. A rendering of such a detector module is depicted in Fig.~\ref{f:module}.

\begin{figure}
   \subfigure[]{
        \includegraphics[width=0.4\columnwidth]{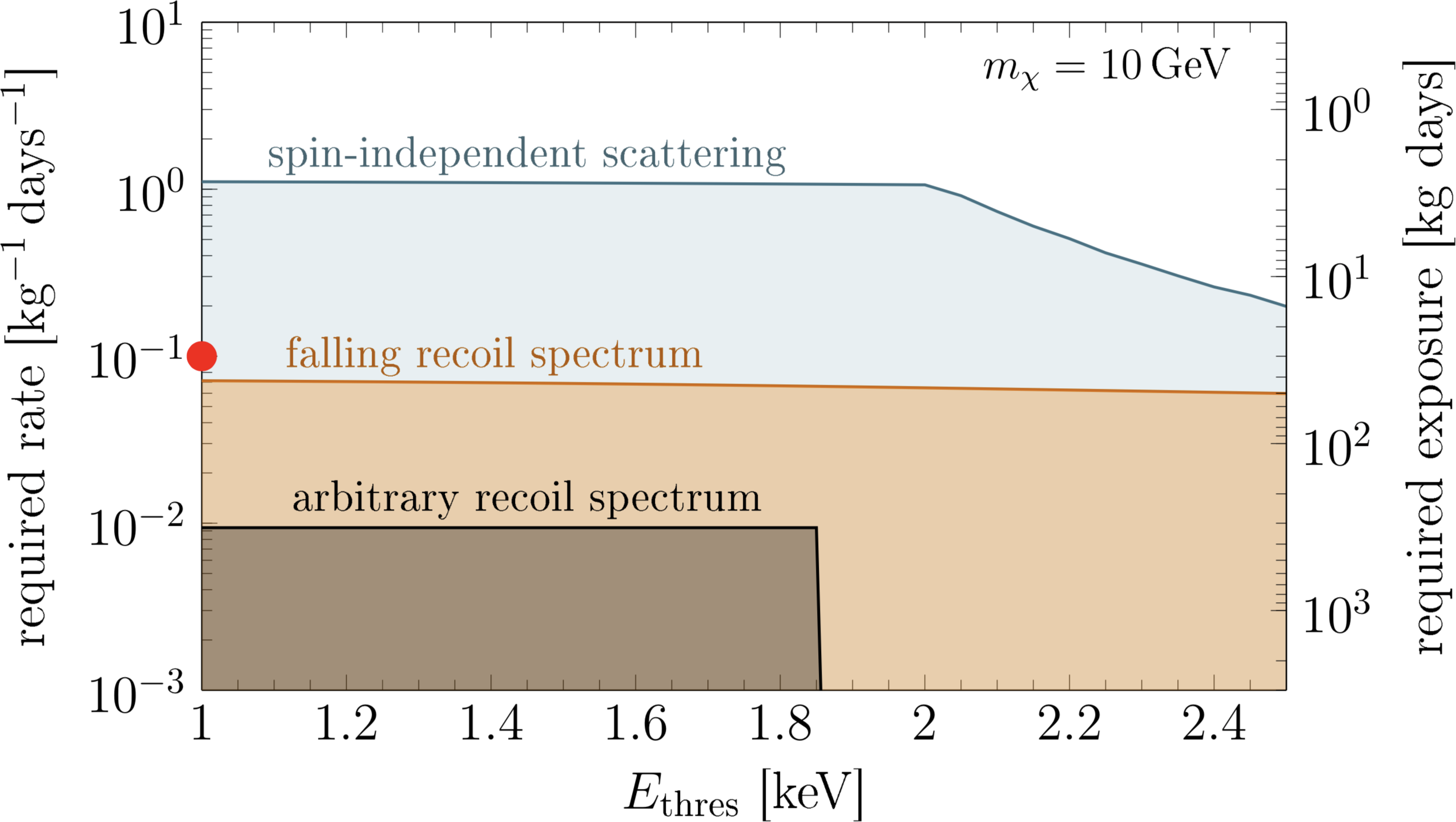}\label{f:physics_reach}}
        \subfigure[]{
        \includegraphics[width=0.59\columnwidth]{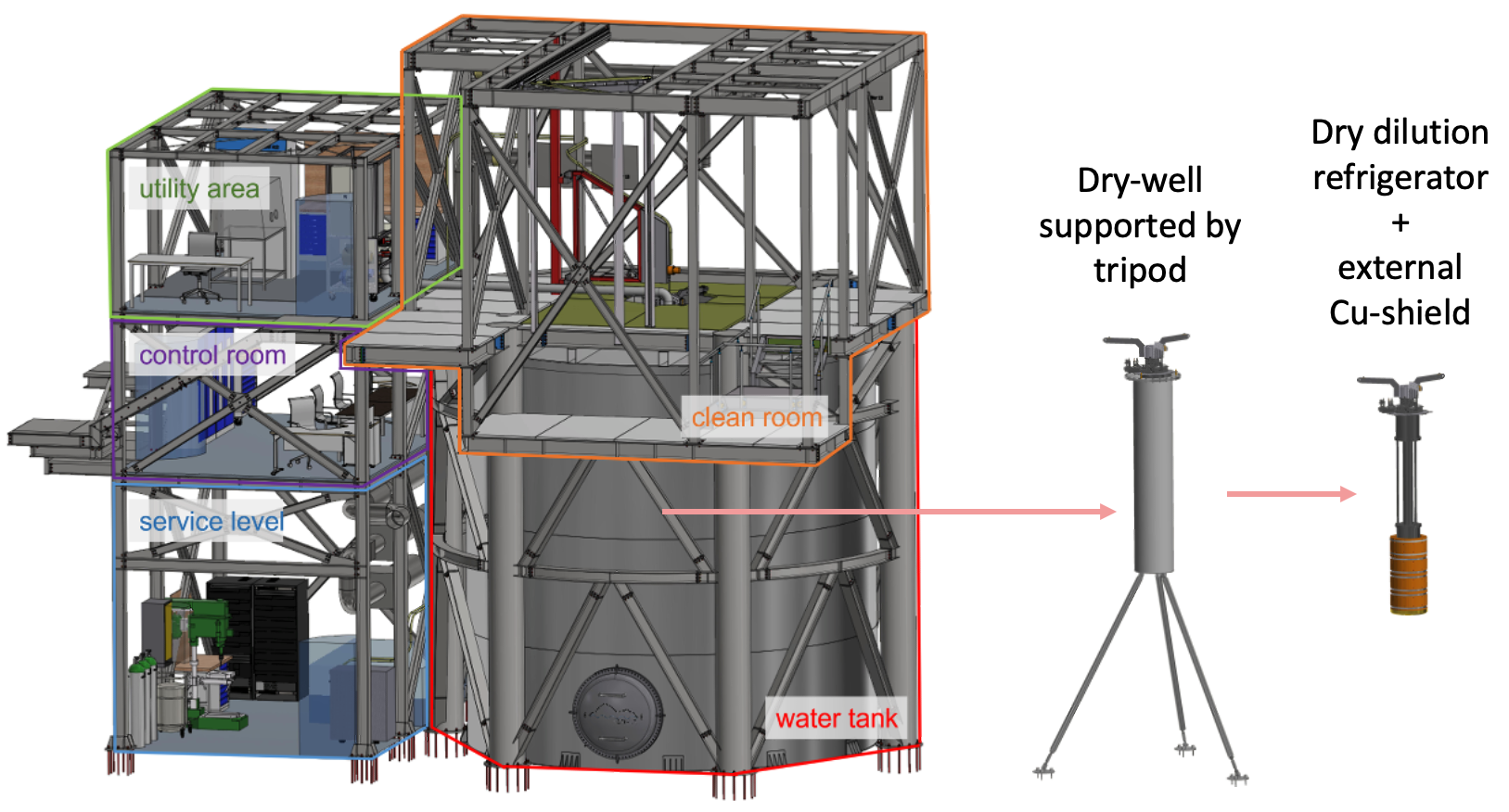}\label{f:facility}}
\caption{ (a)~Required bound on the interaction rate in COSINUS as a function of recoil energy for an arbitrary DM-nucleus interaction (brown), a falling recoil spectrum (light brown) and spin-independent interactions (blue); all three examples are independent of the DM halo. The right y-axis shows the needed exposure assuming a background-free experiment~\cite{kahlhoefer_model-independent_2018}. (b)~COSINUS experimental facility located in hall B of the Gran Sasso Underground Laboratory in Italy: the dry dilution refrigerator hosting the detectors is inserted in a dry-well located in the center of a water tank serving as passive and active muon veto system.}
\label{f:COSINUS_reach}
\end{figure}

Looking in the future reach of COSINUS, Fig.~\ref{f:physics_reach} visualizes the rate (left y-axis) and corresponding exposure (right y-axis) required to exclude a DM interpretation of the DAMA/LIBRA signal (m$_\chi$=\unit[10]{GeV/c$^2$}) in a halo and interaction independent way is shown. The red dot indicates an exposure of \unit[100]{kgd}.  Different colors correspond to different assumptions on the DM recoil spectrum. Thereby, dark brown depicts an exclusion without presuming any model on the DM halo or interaction properties except for the assumption that DM induces nuclear recoils.   If one instead applies stricter assumptions less exposure is necessary: light brown for any recoil spectrum falling with energy and blue for SI scattering (but both still halo-independent). The experimental requirements of COSINUS for a model-independent cross-check of DAMA with a cryogenic NaI-detector are: (i) a nuclear recoil threshold of about \unit[1]{keV}, and (ii) a bound on the interaction rate as low as \unit[10$^{-2}$]{counts/(kg-day)} or a required gross exposure in the ballpark of \unit[1000]{kg-days} before cuts.

\subsubsection{The low-background cryogenic facility}
The COSINUS experiment is hosted by the Laboratori Nazionali del Gran Sasso. Following~\cite{angloher2021simulationbased}, the COSINUS underground low-background facility, see Fig.~\ref{f:facility}, consists of four main parts: i) the dilution refrigerator, located in a dry-well inside the water tank, provides the low temperatures necessary to operate the detectors at milliKelvin temperatures, ii) the water tank shields ambient radioactivity, iii) a servicing level equipped as clean room allows to mount detectors and includes a lifting system to move the cryostat in and out of the dry-well and the external Cu shield and iv) a two-floor control building next to the tank.

Since the cosmogenic neutron background is up to two orders of magnitude higher than the radiogenic one, the use of an active veto system is needed. Indeed Monte Carlo simulations show that the rate of cosmogenic neutron events is in the order of $(3.5 \pm 0.7)$~cts kg$^{-1}$ yr$^{-1}$~\cite{angloher2021simulationbased} which corresponds to roughly one event in \unit[100]{kg-days}. With an active muon veto, consisting of 30 photo multiplier tubes installed on the mantle and bottom of the water tank, we anticipate reducing the rate of cosmogenic neutrons down to the one of radiogenic neutrons which the Monte Carlo simulation predicts to be 0.05~cts kg$^{-1}$ yr$^{-1}$.  In order to achieve a complete background model of COSINUS these simulations will be further refined once the geometry of the experiment is fully fixed in all details. In addition, radiopurity screening efforts are going on at the moment for all materials in the vicinity of the detectors.

The facility construction started in 2020 and is now at the last stage, the commissioning of the experimental infrastructure, particularly the dry dilution refrigerator. 
First data taking with eight NaI-detectors of \unit[$\sim$30]{g} each is anticipated to start early 2025. The complete model-independent clarification of a nuclear recoil origin of the DAMA signal is sought for within five years once the required detector performance of \unit[1]{keV} energy threshold for a massive NaI crystal of $\mathcal{O}$(\unit[100]{g}) is achieved. The necessary exposure goal is about \unit[1000]{kg-days} and using a total of 24 detectors~\cite{kahlhoefer_model-independent_2018}.

\section{{Conclusions and outlook}}

To conclude, the two authors, members of the CRESST and COSINUS collaborations, will draw a summary with a personal touch, and only time will tell as to their accuracy. Identifying the nature of dark matter is among the highest priority topics of modern astroparticle physics. 
Direct searches play a crucial role in determining the nature of dark matter particles and possible new fundamental forces. Experiments employing advanced low-temperature detectors with extremely low eV-scale energy thresholds have the potential to lead to ground-breaking discoveries in the field of rare-event searches. 

Low-temperature detectors have demonstrated in the past 35 years to be very flexible and versatile, adapting their performance from massive kg-scale calorimeters employed for the hunt of the classical WIMPs to now extremely sensitive low-threshold detectors for exploring light dark matter particles with unprecedented sensitivities. In fact, phonon detectors employing advanced superconducting sensors like TESs are naturally excellent ultra-low threshold detectors as they measure the mean energy of all excitations, hereby maximizing the signal and enabling ultimate sensitivities. 

However, the sensitivity of all solid-state low-threshold cryogenic experiments is currently limited by new challenges related to detector physics artifacts that arise at energies below several hundred eV, known as low-energy excesses. At the moment, the low-temperature detector community is focused on getting an understanding of the origin of the LEE. Besides the LEE, there are other challenges associated with entering this new energy regime, such as the need for new calibration methods and rigorously controlled backgrounds. Also the fabrication of larger arrays of cryogenic detectors with reproducible performance is not yet in reach. 

A novel approach employing superfluid helium as a target for DM search is currently undergoing active development by several new collaborations. A superfluid target is expected to eliminate lattice- and stress-related processes and offers straightforward scalability. Employing advanced superconducting sensors to read out signals from helium evaporation and scintillation presents another potential application for low-temperature detector technologies at cutting-edge science.

CRESST employs scintillating cryogenic calorimeters and is currently one of the experiments pushing the light dark matter frontier through excellent detector performance. Simultaneous readout of produced scintillation light enables particle discrimination at the keV-energy scale and provides valuable input for background simulations.
The COSINUS experiment, a spin-off of CRESST, adapted the TES-readout technology ("remoTES") to delicate absorbers like sodium iodide. The phonon-light technique allows to identify nuclear recoils from electron recoils in the NaI detector enabling a fully model-independent cross-check of the DAMA/LIBRA signal. This validation is essential to solve this long-lasting and puzzling situation that keeps the direct DM search community disunited.

The versatility and flexibility of low-temperature detectors have been and continue to be key features that open up new physics applications. Adapting this technology to other fields seems to be a natural development. Fundamental neutrino physics studies in the frame of CE$\nu$NS experiments or neutrino mass measurements have already indicated new directions.
Over the next decade, progress is expected to center on the improvement of detector technologies and a deeper understanding of signal formation with the near goal of probing vast regions of well-motivated but yet unexplored parameter space at low dark matter particle masses.

\section*{Acknowledgements}
We are grateful to the CRESST and COSINUS collaborations for providing the materials and insights for this review. The sensitivity projections for CRESST were calculated with the CRESST/COSINUS collaboration-internal software tool \textit{limitless}. We would like to thank Leonie Einfalt (HEPHY, TU Wien) for her generous support with the projections calculation and data preparation, Florian Reindl (HEPHY, TU Wien) for support with the data used in this work and his feedback on the manuscript, as well as Moritz Kellermann (MPP, Garching), Robert Stadler (MPP, Garching), and Karlheinz Ackermann (MPP, Garching) for their valuable inputs to the included schematics. We thank Belina von Krosigk (Heidelberg University), Julien Billard (IPNL, Lyon), Mariano Cababie (HEPHY, TU Wien), and Raimund Strauss (TU Munich) for their inputs to this work, and Franz Pröbst (MPP, Garching) and Maximilian Hughes (MPP, Garching) for proof-reading the manuscript.

\bibliographystyle{unsrt2}
\bibliography{ref2}

\end{document}